\documentclass[preprint]{vldb}
\usepackage{etex}
\usepackage{booktabs} 
\usepackage{balance}
\usepackage[T1]{fontenc}
\usepackage{hyperref}
\usepackage[linesnumbered,ruled]{algorithm2e}
\usepackage{paralist}
\usepackage{multirow}
\usepackage{xspace}
\usepackage{url}
\usepackage{float}
\usepackage{subfig}
\usepackage{times}
\usepackage{soul}
\usepackage{tabularx}
\usepackage{algpseudocode}
\usepackage{xcolor}
\usepackage{color}
\usepackage{tikz}
\usepackage{pgfplots}
\usepackage{listings}
\usepackage{caption}
\usepackage{amsmath,amssymb,amsfonts,graphicx,nicefrac,mathtools}
\usepackage{thmtools,thm-restate}
\usepackage{enumitem}
\usepackage{grffile}
\usepackage[normalem]{ulem}
\usepackage{tcolorbox}
\usepackage[framemethod=TikZ]{mdframed}
\usepackage{lipsum}
\usepackage{array}
\usepackage{cleveref}

\usetikzlibrary{arrows,decorations.pathreplacing}

\newcolumntype{C}[1]{>{\centering\arraybackslash}m{#1}}
\newcolumntype{M}[1]{>{\arraybackslash}m{#1}}

\input{vldbreffix.tex}

\newenvironment{algo}{%
  \algorithm
}{%
  \endalgorithm
}

\xspaceaddexceptions{[]\}(}

\allowdisplaybreaks

\pgfplotsset{
  width=6cm,
  height=4cm, 
  compat=newest,
  xlabel near ticks,
  ylabel near ticks}

  \mdfdefinestyle{MyFrame}{%
    linecolor=gray,
    outerlinewidth=0.25pt,
    roundcorner=3pt,
    innertopmargin=\baselineskip,
    innerbottommargin=\baselineskip,
    innerrightmargin=10pt,
    innerleftmargin=10pt,
    backgroundcolor=white}

\newcommand{\hidden}[1]{}

\newcommand{\papertext}[1]{\ignorespaces}
\newcommand{\techreport}[1]{#1}

\makeatletter
\def\@copyrightspace{\relax}
\makeatother

\newcommand{\BIT}{\begin{itemize}}
\newcommand{\EIT}{\end{itemize}}
\newcommand{\BNUM}{\begin{enumerate}}
\newcommand{\ENUM}{\end{enumerate}}
 
\def\reals{\mathbb{R}} 
\def\bigo#1{\mathcal{O}\left(#1\right)} 
\renewcommand{\exp}[1]{\operatorname{exp}\left(#1\right)} 

\def\E{\mathbb{E}} 
\def\Earg#1{\E\left[{#1}\right]}

\def\P{\mathbb{P}} 
\def\Parg#1{\P\left({#1}\right)}
\def\Psubarg#1#2{\P_{#1}\left[{#2}\right]}

\def\HypGeo{\textnormal{HypGeo}}

\newcommand{\veps}{\varepsilon}

\newcommand{\fst}{^{(1)}}
\newcommand{\snd}{^{(2)}}
\newcommand{\nth}{^{(n)}}

\newcommand{\ith}{^{(i)}}
\newcommand{\jth}{^{(j)}}

\newcommand{\tth}{^{(t)}}

\def\norm#1{||#1||}

\def\minspeedup{8\times\xspace}
\def\maxspeedup{35\times\xspace}
\def\viztype{histogram\xspace}
\def\viztypes{histograms\xspace}

\soulregister{\viztype}{7}
\soulregister{\viztypes}{7}

\newcommand{\vzero}{\mathbf{0}}

\newcommand{\vx}{{V_X}}
\newcommand{\vz}{{V_Z}}
\newcommand{\mvx}{{|\vx|}}
\newcommand{\mvz}{{|\vz|}}
\newcommand{\vQ}{\mathbf{q}}
\newcommand{\vhQ}{\mathbf{\bar{q}}}

\newcommand{\mbr}{\mathbf{\bar{r}}}
\newcommand{\bfr}{\mathbf{r}}
\newcommand{\brr}{\bar{r}}

\newcommand{\vr}{\bfr}
\newcommand{\vrs}{\bfr^*}
\newcommand{\vri}{\bfr_i}
\newcommand{\vrip}{\vri^\partial}
\newcommand{\vrj}{\bfr_j}

\newcommand{\vrsi}{\bfr_i^*}
\newcommand{\vrsj}{\bfr_j^*}
\newcommand{\vhr}{\mbr}
\newcommand{\vhri}{\vhr_i}
\newcommand{\vhrip}{\vhr_i^\partial}

\newcommand{\vhrs}{\vhr^*}
\newcommand{\vhrsi}{\vhrs_i}

\newcommand{\hrj}{\brr_j}
\newcommand{\hrsj}{\brr^*_j}

\newcommand{\taui}{\tau_i}
\newcommand{\tauj}{\tau_j}
\newcommand{\tausi}{\tau_i^*}
\newcommand{\tausj}{\tau_j^*}
\newcommand{\tauip}{\tau_i^\partial}
\newcommand{\taujp}{\tau_j^\partial}

\def\lookahead{\texttt{lookahead}\xspace}

\def\gsep{\Cref{guarantee:separation}\xspace}
\def\grec{\Cref{guarantee:reconstruction}\xspace}
\def\guarantees{\Cref{guarantee:separation,guarantee:reconstruction}\xspace}

\def\Ntotal{N}
\def\Ntotali{N_i}
\def\Ntotalj{N_j}
\def\takenfori{n_i}
\def\takenforip{n_i^\partial}

\def\totakefori{n_i'}

\def\minsel{\sigma}

\newcommand{\matchingset}{M}
\newcommand{\activeset}{A}
\newcommand{\nonmatchingset}{\activeset\setminus\matchingset}

\newcommand{\matchingbound}{s-\frac{\veps}{2}}
\newcommand{\nonmatchingbound}{s+\frac{\veps}{2}}

\newcommand{\vepsi}{$\veps_i$\xspace}

\newcommand{\deli}{$\delta_i$\xspace}
\newcommand{\deltaupper}{\delta^{upper}}

\newcommand{\numflightstuples}{606 million\xspace}
\newcommand{\preck}{$k$\xspace}

\newcommand{\flights}{\textsc{Flights}\xspace}

\newcommand{\taxi}{\textsc{Taxi}\xspace}
\newcommand{\police}{\textsc{Police}\xspace}

\def\scan{{\sf Scan}\xspace}
\def\slowmatch{\red{\sf SlowMatch FIXME}\xspace}
\def\scanmatch{{\sf ScanMatch}\xspace}
\def\syncmatch{{\sf SyncMatch}\xspace}

\def\anyactive{{\sf AnyActive}\xspace}

\newcommand\SUM{\textsc{SUM}\xspace}
\newcommand\COUNT{\textsc{COUNT}\xspace}

\newtheorem{theorem}{Theorem}
\newtheorem{lemma}{Lemma}

\newtheorem{definition}{Definition}
\newtheorem{guarantee}{Guarantee}
\crefname{guarantee}{Guarantee}{Guarantees}
\newtheorem{problem}{\textbf{Problem}}
\crefname{problem}{Problem}{Problem}

\newcommand{\red}[1]{#1}
\newcommand{\blue}[1]{#1}

\newcommand{\agp}[1]{\ignorespaces}
\newcommand{\agpres}[1]{\ignorespaces}
\newcommand{\smacke}[1]{\ignorespaces}
\newcommand{\smackeres}[1]{\ignorespaces}
\newcommand{\silu}[1]{\ignorespaces}

\newcommand{\smackout}[2]{\red{#2}}
\newcommand{\smackoutres}[2]{\red{#2}}
\newcommand{\resolved}[1]{\ignorespaces}
\newcommand{\resolvedres}[1]{\ignorespaces}
\newcommand{\strike}[1]{\ignorespaces}

\newcommand{\zv}{{\sf zenvisage}\xspace}

\newcommand{\fm}{{\sf FastMatch}\xspace}
\newcommand{\vs}{{\sf HistSim}\xspace}
\newcommand{\hsim}{\vs}

\newcommand{\histsim}{\vs}
\newcommand{\algname}{\fm}

\newenvironment{denselist}{
    \begin{list}{\small{$\bullet$}}%
    {\setlength{\itemsep}{0ex} \setlength{\topsep}{0ex}
    \setlength{\parsep}{0pt} \setlength{\itemindent}{0pt}
    \setlength{\leftmargin}{1.5em}
    \setlength{\partopsep}{0pt}}}%
    {\end{list}}

\newcommand{\topic}[1]{\vspace{-3.5pt}\smallskip \smallskip \noindent{\bf #1.}}
\newcommand{\stitle}[1]{\vspace{0.5em}\noindent\textbf{#1}}
\newcommand{\emtitle}[1]{\vspace{0.3em}\noindent{\em #1}}
\newcommand{\frameme}[1]{

\noindent\fbox{
  \parbox{0.95\linewidth}{
    \noindent #1
    }
  }

}

\newcommand{\figs}{./}

\numberofauthors{1}

\begin{document}

\def\com#1{\emtitle{#1}}

\newcommand{\authdag}{$^\dagger$}
\newcommand{\authddag}{$^\ddagger$}

%
%

\title{Adaptive Sampling for Rapidly Matching Histograms}

\author{
\alignauthor Stephen Macke\authdag, Yiming Zhang\authddag, Silu Huang\authdag, Aditya Parameswaran\authdag \\
       \affaddr{University of Illinois-Urbana Champaign} \\
       \affaddr{\authdag\{smacke,shuang86,adityagp\}@illinois.edu | \authddag ym\_zhang@sjtu.edu.cn}
}


%

\toappear{
Permission to make digital or hard copies of all or part of this work
for personal or classroom use is granted without fee provided that copies are not
made or distributed for profit or commercial advantage and that copies bear this
notice and the full citation on the first page. To copy otherwise, to republish,
to post on servers or to redistribute to lists, requires prior
specific permission and/or a fee.\\
Copyright 20XX ACM X-XXXXX-XX-X/XX/XX
...\$15.00.
}

\date{\today}

\maketitle

\begin{abstract}

In exploratory data analysis, analysts often have a need
to identify \viztypes that possess a specific distribution,
among a large class of candidate \viztypes, e.g., 
find countries whose income 
distribution is most similar to that of Greece.
This distribution could be a new one that 
the user is curious about, 
or a known distribution from an existing \viztype visualization.
At present, this process of 
identification is brute-force,
requiring the manual generation and 
evaluation of a large number of \viztypes.
We present \algname: an end-to-end approach for 
interactively retrieving
the \viztype visualizations most similar to a
user-specified target,
from a large collection of \viztypes.
The primary technical contribution underlying \algname is
a \red{probabilistic} algorithm, \hsim, a theoretically sound
sampling-based approach to 
identify the top-$k$ closest \viztypes under
$\ell_1$ distance.
While \hsim can  be used independently, 
within \fm we couple \hsim with a novel system
architecture that is aware of practical considerations,
employing \red{asynchronous} block-based sampling policies,
building on lightweight sampling engines developed
in recent work~\cite{kim2014needletail}.
\algname obtains near-perfect accuracy with up
to \blue{$\maxspeedup$}
speedup over \blue{approaches that do not use sampling}
on several real-world datasets.

\end{abstract}

\section{Introduction}
\label{sec:intro}

In exploratory data analysis, 
analysts often generate 
and peruse a large number
of visualizations to identify those that 
{\em match desired criteria}.  
This process of iterative ``generate and test'' 
occupies a large part of visual data
analysis~\cite{behrens1997principles,Hanrahan:2012:ADT:2213836.2213902,pirolli2005sensemaking}, 
and is often cumbersome and time consuming, especially on very large
datasets that are increasingly the norm.
This process ends up impeding interaction, preventing exploration, and delaying
the extraction of insights.

\techreport{\emtitle{Example 1: Census Data Exploration. }}\papertext{\emtitle{Motivating Example: Census Data Exploration. }}
 Alice is exploring a census dataset
 consisting of hundreds of millions of tuples,
 with attributes such as gender, occupation, nationality, ethnicity,
 religion, adjusted income, net assets, 
 and so on. 
 In particular, she is interested in understanding how applying various
 filters impacts the relative distribution of tuples with
 different attribute values.
 She might ask questions like
 {\em Q1:} Which countries have similar distributions of wealth to that of Greece? 
 {\em Q2:} In the United States, which professions have an ethnicity
 distribution similar to the profession of doctor?
 {\em Q3:} Which (nationality, religion) pairs have a similar distribution
 of number of children to Christian families in France?

\techreport{\emtitle{Example 2: Taxi Data Exploration.}
Bob is exploring the distribution of taxi trip times originating from various
locations around Manhattan. Specifically, he
plots a \viztype showing the distribution of taxi pickup times
for trips originating from various locations. 
As he varies the location, he examines how
the \viztype changes, and he notices that choosing the location
of a popular nightclub skews the distribution of pickup times heavily in the range
of 3am to 5am. He wonders {\em Q4:} Where are the other locations around Manhattan
 that have similar distributions of pickup times? {\em Q5:} Do they all
 have nightclubs, or are there different reasons for the late-night pickups?

\emtitle{Example 3: Sales Data Exploration.}
 Carol has the complete history of all sales at a large
 online shopping website. Since users must enter birthdays in order to create
 accounts, she is able to plot the age distribution of purchasers for any
 given product. To enhance the website's recommendation engine,
 she is considering recommending products with similar purchaser
 age distributions. To test the merit of this idea, she first wishes to perform
 a series of queries of the form {\em Q6:}
 Which products were purchased by users with ages most closely following the
 distribution for a certain product---a particular
 brand of shoes, or a particular book, for example?
 Carol wishes to
 perform this query for a few test
 products before integrating this feature into the recommendation pipeline.}
 
\smallskip
\noindent
\techreport{These cases represent scenarios that often arise }\papertext{This example represents a scenario that often arises }
in exploratory data analysis---finding matches to a specific distribution.
The focus of this paper is to \ul{\em develop techniques
for rapidly exploring a large class of \viztypes{} to find those
that match a user-specified target}.

\techreport{Referring to {\em Q1} in the first example,}\papertext{Referring to {\em Q1} in our motivating example,}
a typical workflow used by Alice may be the following:
first, pick a country. Generate the corresponding \viztype.
\techreport{This could be done either using a language like R, Python, or Javascript, with the 
visualization generated
in ggplot~\cite{wickham2016ggplot2} or D3~\cite{bostock2011d3},
or using interactions in a visualization platform like Tableau~\cite{stolte2002polaris}.} 
Does the visualization look similar to that of Greece?  
If not, pick another, generate it, and repeat.
Else, record it, pick another, generate it, and repeat.  
If only a select few countries
have similar distributions, she may spend a huge amount of time 
sifting through her
data, or may simply give up early.

\stitle{The Need for Approximation.}
Even if Alice generates all of the 
candidate \viztypes{} (i.e., one for each country) in a single pass,
programmatically selecting
the closest match to her target 
(i.e., the Greece \viztype), 
this could take unacceptably long.
If the dataset is tens of gigabytes and
every tuple in her census dataset 
contributes to some \viztype, then any exact
method must necessarily process tens
of gigabytes---on a typical
workstation, this can take
tens of seconds
even for in-memory data.
Recent work suggests that 
latencies greater than 500ms cause significant
frustration for end-users and lead them to test fewer hypotheses
and potentially identify fewer insights~\cite{liu2014effects}.
Thus, in this work, we explore approximate techniques
that can return matching \viztype visualizations
with accuracy guarantees, but much faster. 

One tempting approach is to employ approximation
using pre-computed samples~\cite{blinkdb,acharya1999aqua,acharya2000congressional,Babcock2003,icicles,SPS}, 
or pre-computed sketches or other summaries~\cite{wavelets,Park2015,wu2015efficient}.
Unfortunately, in an interactive exploration setting, pre-computed
samples or summaries are not helpful, since
the workload is unpredictable and changes rapidly,
with more than half of the queries issued one week completely absent in the following week,
and more than 90\% of the queries issued one week completely absent a month later~\cite{mozafari2017approximate}.
In our case, based on the results for one matching query, Alice 
may be prompted to explore different (and arbitrary) slices of the same data,
which can be exponential in the number of attributes in the dataset.
Instead, we \blue{materialize samples on-the-fly}, which doesn't suffer from the same limitations
and has been employed for generating approximate visualizations incrementally~\cite{rahman2016ve}, and 
while preserving ordering and perceptual guarantees~\cite{Kim2015,alabi2016pfunk}.
To the best of our knowledge, however, \blue{on-demand approximate sampling}
techniques have not been applied to the problem
of evaluating a large number of visualizations for matches in parallel.


\stitle{Key Research Challenges.}
In developing an approximation-based approach  
for rapid \viztype matching 
we immediately 
encounter a number of theoretical and practical 
challenges.

\emtitle{1. Quantifying Importance.} 
To benefit from approximation,
we need to be able to quantify which 
samples are ``important'' to facilitate progress
towards termination. 
It is not clear how to assess this importance:
at one extreme, it may be preferable to sample more
from candidate \viztypes that are more ``uncertain'',
but these \viztypes may already be known to be rather 
far away from the target. 
Another approach is to sample more from candidate
\viztypes at the ``boundary'' of top-$k$, but if these
\viztypes are more ``certain'', refining them
further may be useless.
Another challenge is when we quantify the importance
of samples: one approach would be to reassess importance every time 
new data become available, but this approach could be
computationally costly.
\smackoutres{or we delay assessment of importance to after every $m$ samples,
but this may lead to a stale estimate.}{}
\agpres{why delete this?}\smackeres{now that the
algorithm is divided into rounds with well-defined boundaries, it's a bit clearer
when the importance quantification should be performed (once per round).
Less clear is how to divide up the rounds; I identify that as the
major challenge of the next point.}

\emtitle{2. Deciding to Terminate.}
Our algorithm needs to ascribe a confidence in the correctness of 
partial results in order to determine when it
may safely terminate.  
This ``confidence quantification'' requires
\red{performing a statistical test. If we perform
this test too often, we spend a significant amount of time doing computation
that could be spent performing I/O, and we further lose statistical power since
we are performing more tests; if we do not do this test often enough, we may
end up taking many more samples than are necessary to terminate.}

\emtitle{3. Challenges with Storage Media.}
When performing sampling from traditional storage media, 
the cost to fetch samples is locality-dependent;
truly random sampling is extremely expensive due to random I/O, 
while sampling at the level of blocks is much more efficient, 
but is less random.

\emtitle{4. Communication between Components.}
It is crucial for our overall system 
to not be bottlenecked on any \smackoutres{component:
in quantifying importance (via the sampling manager),
deciding whether to terminate (via the statistics manager),
and retrieving samples (via the I/O manager). 
Without this, the time taken for execution can often
be greater than the time taken by exact methods.}{component.
In particular, the process of quantifying importance (via the sampling
manager) must not block the actual I/O performed; otherwise,
the time for execution may end up being greater than the
time taken by exact methods.}
As such, these components must proceed asynchronously,
while also minimizing communication
across them.


\stitle{Our Contributions.}
In this paper, we have
developed an end-to-end architecture for \viztype
matching, dubbed \fm, addressing the challenges identified above:

\hidden{\emtitle{0. Distance Computation for a Single Histogram.}
 \smackout{We prove an information-theoretically 
 optimal upper bound on the number
 of samples needed for an 
 empirically-learned discrete distribution to be
 $\veps$-close (under $\ell_1$ distance)  
 to the true discrete distribution from
 which it was sampled. 
 Our \fm system uses this guarantee to
 obtain bounds on the distance of a distribution 
 from an analyst-provided target.
 In particular, this bound is tighter than, and therefore more useful
 than those derived in prior work in the theory literature.
 (We discuss this further in Section~\ref{subsec:confidence}.)}{}
 \smacke{Getting rid of this since it is ``known'' to theory people.}
 \agp{make sure you update this to reflect changes. 1-3, and outline}}

\emtitle{1. Importance Quantification Policies.}
We develop a sampling engine that employs a simple
and theoretically well-motivated criterion
for deciding whether processing particular
portions of data will allow for
faster termination.
Since the criterion is simple,
it is {\em easy to update} as we process new data,
``understanding'' when
it has seen enough data for some \viztype,
or when it needs to take more data to distinguish \viztypes that are
close to each other.

\emtitle{2. Termination Algorithm.}
We develop a statistics engine that
repeatedly performs a lightweight ``safe termination'' test,
based on the idea of \smackoutres{relaxing fixed-width confidence intervals
into variable-width deviation bounds}{performing multiple hypothesis tests
for which simultaneous rejection implies correctness of the results}.
\smackoutres{This relaxation allows these
deviation bounds to borrow statistical strength from each other,
minimizing the time necessary for safe termination.
Since this test is extremely efficient (asymptotically
dominated by a distance computation for each \viztype),
the statistics engine is able to run this frequently enough
to ensure timely termination.}{Our statistics engine further
quantifies how often to run this test to ensure timely termination
without sacrificing too much statistical power.}
\agpres{I understand this but I am not sure how it is implied by your following statement.}
\agpres{I am not sure DB reviewers will ``get'' the notion of hyp tests and statistical power.
If possible introduce this in a less technical way.}
\smackeres{\textbf{Does this make sense? We want the test to run often enough so that
if it is in a state where it will reject, we will perform the test ASAP
so as to avoid wasting time taking more samples, but every time we take the
test, we halve the level (i.e. $\delta/4$, $\delta/8$, $\delta/16$, and so on),
so that too-short-of-time between tests will greatly increase the amount of data
we need to read to get the test to reject.}}

\emtitle{3. Locality-aware Sampling.}
To better exploit locality of storage media,
\fm samples at the level of blocks, proceeding sequentially.
To estimate the benefit of blocks, we \smackoutres{develop a
just-in-time lookahead technique for deciding that blocks to read,}{leverage bitmap indexes}
in a cache-conscious manner, evaluating \smackoutres{the blocks}{multiple blocks at a time}
in the same order as their layout in storage.
Our technique minimizes the time
required for the query output to satisfy our probabilistic guarantees.

\emtitle{4. Decoupling Components.}
Our system decouples the overhead of deciding which samples to take from the
actual I/O used to read the samples from storage. \red{In particular, our sampling
engine utilizes a just-in-time lookahead technique that marks blocks for reading
or skipping while the I/O proceeds unhindered, in parallel.}
\smackoutres{As such, the computation cannot block I/O.
Furthermore, we minimize communication overhead introduced by this decoupling using
some smart but simple optimizations that exploit the fact that we are dealing with
 \viztypes.}{}


\smallskip
\noindent
Overall, we implement \fm within the context of a bitmap-based
sampling engine, which allows us to quickly determine whether a given \red{memory or} disk
block \smackoutres{contains}{could contain} samples matching ad-hoc predicates. Such
engines were found to effectively support approximate generation
of visualizations in recent work~\cite{alabi2016pfunk,Kim2015,rahman2016ve}. 

\smallskip
\noindent
We find that our approximation-based techniques working in tandem with our
novel systems components {\bf \em  lead to speedups ranging from
$\minspeedup$ to over $\maxspeedup$}
over exact methods, and moreover, unlike less-sophisticated variants of \fm, 
whose performance can be highly data-dependent,
\fm consistently brings latency to near-interactive levels.

\stitle{Related Work.}
To the best of our knowledge, 
there has been no work on sampling to identify
\viztypes that match user specifications. 
Sampling-based techniques have been applied to generate
visualizations that preserve visual properties~\cite{alabi2016pfunk,Kim2015},
and for incremental generation of time-series and heat-maps~\cite{rahman2016ve}---all focusing
on the generation of a single visualization. 
Similarly, Pangloss~\cite{moritz2017trust} employs approximation via the Sample+Seek approach~\cite{SPS}
to generate a single visualization early, while minimizing error.
\techreport{\blue{One system uses
workload-aware indexes called ``VisTrees''~\cite{tim2016vistrees}
to facilitate sampling for interactive generation of histograms {\em without} error guarantees.}}
M4 uses rasterization without sampling to reduce the dimensionality of a time-series
visualization and generate it faster~\cite{jugel2014m4}.
SeeDB~\cite{vartak2015s} recommends visualizations
to help distinguish between two subsets of data
while employing approximation.
\smackoutres{However, their
objectives are different, and their approach does not provide
any guarantees.}{However, their techniques are tailored to evaluating
differences between pairs of visualizations (that share the same axes,
while other pairs do not share the same axes). In our case, we need to compare
one visualization versus others, all of which have the same axes and have
comparable distances, hence the techniques do not generalize.}

Recent work has developed 
\zv~\cite{siddiqui2016effortless}, a visual exploration tool, including
operations that identify visualizations
similar to a target.
However, to identify matches, \zv does not consider sampling,
and requires at least
one complete pass through the dataset.
\fm  was developed as a back-end with such interfaces in mind 
to support rapid discovery of relevant visualizations.

\papertext{Additional related work is surveyed in \Cref{sec:related}.}

\stitle{Outline.} 
\Cref{sec:problem}
articulates the formal problem of identifying top-$k$ closest \viztypes to
a target. \Cref{sec:algorithm} introduces our \hsim
algorithm for solving this problem, while \Cref{sec:arch}
describes the system architecture that implements this algorithm.
In \Cref{sec:experiments} we perform an empirical evaluation
on several real-world
datasets.
\papertext{Several generalizations and extensions are omitted from the paper
and can be found in our extended technical report~\cite{techreport}.}
\techreport{After surveying additional related work in \Cref{sec:related},
we describe several generalizations and extensions of our techniques
in \Cref{sec:extensions}.}


\pgfplotstableread{
X Y
1 25
2 15
3 65
4 45
5 60
6 20
7 23
}\querytable

\pgfplotstableread{
X Y  Yerr
1 15 5
2 20 10
3 60 3
4 47 8
5 45 5
6 30 10
7 23 3
}\candtable

\begin{figure}[t]
  \centering
\subfloat{\scalebox{0.8}{
\begin{tikzpicture}[thick,scale=1.0,every node/.style={xscale=1.9}]
\tikzstyle{every node}=[font=\small]
\begin{axis}[
    title=Target Histogram (Greece),
    ybar,
    bar width=15pt,
    bar shift=0pt,
    xlabel = {Income Bracket},
    ylabel = {Population Counts},
    symbolic x coords = {1,2,3,4,5,6,7},
    ymin=0,
    ytick=\empty,
    xtick = data,
    axis x line=bottom,
    axis y line=left,
    enlarge x limits=0.15,
    xticklabel style={anchor=base,yshift=-\baselineskip},
    ]
    \addplot+[
      ] table[x=X,y=Y] {\querytable};
\end{axis}
\end{tikzpicture}
}}
\subfloat{\scalebox{0.8}{
\begin{tikzpicture}[thick,scale=1.0,every node/.style={xscale=1.9}]
\tikzstyle{every node}=[font=\small]
\begin{axis}[
    title=Candidate (\texttt{\$Country=Italy}),
    ybar,
    bar width=15pt,
    bar shift=0pt,
    xlabel = {Income Bracket},
    ylabel = {Population Counts},
    symbolic x coords = {1,2,3,4,5,6,7},
    ymin=0,
    ytick=\empty,
    xtick = data,
    axis x line=bottom,
    axis y line=left,
    enlarge x limits=0.15,
    xticklabel style={anchor=base,yshift=-\baselineskip},
    ]
    \addplot+[
        color=red,
        fill=pink,
        error bars/.cd,
        y dir=both,
        y explicit,
      ] table[x=X,y=Y] {\candtable};
\end{axis}
\end{tikzpicture}
}}

\vspace{-7pt}
\caption{Example visual target and candidate histogram}\label{fig:interface}
\vspace{-2pt}
\end{figure}
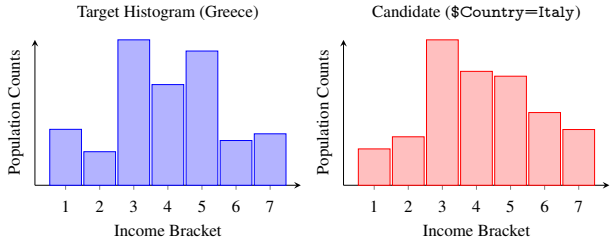

\def\densesection#1{\vspace{-3pt}\section{#1}\vspace{-3pt}}

\densesection{Problem Formulation}
\label{sec:problem}

\begin{table*}[t]
  \centering
  \vspace{-10pt}
  
  \scriptsize
  \begin{tabular}{|C{2.9cm}|M{14.2cm}|}
    \hline
    {\bf Symbol(s)} & {\bf Description} \\ \hline
    $X, Z, \vx, \vz, T$ & x-axis attribute, candidate attribute, respective value sets, and relation over these
    attributes, used in \viztype-generating queries (see \Cref{def:vizquery}) \\ \hline
    $k$, $\delta, \veps, \red{\minsel}$ & User-supplied parameters (number of matching \viztypes to retrieve,
                           error probability upper bound, approximation error upper bound, \red{selectivity threshold
                           (below which candidates may optionally be ignored)} \\ \hline
    $\vQ, \vri, \vrsi, \left(\vhQ, \vhri, \vhrsi\right)$ & Visual target, candidate $i$'s estimated (unstarred) and true (starred) \viztype counts (normalized variants) \\ \hline
    $d(\cdot, \cdot)$ & Distance function, used to quantify visual distance
    (see \Cref{def:cand_distance}) \\ \hline
    $\takenfori$, $\totakefori$, $\veps_i$, $\delta_i$, $\tau_i$ ($\tau_i^*$) &
    Quantities specific to candidate $i$ during \hsim run: number of samples taken, estimated samples needed (see \Cref{sec:arch}), deviation bound (see \Cref{def:deviance}),
    confidence upper bound on \vepsi-deviation \red{or rareness},
    and distance estimate from $\vQ$ (true distance from $\vQ$),
    respectively \\ \hline
    \blue{$\takenforip$, $\vrip$, $\tauip$} &
    \blue{Quantities corresponding to samples taken in a specific round
    of \hsim stage 2: number of samples taken for candidate $i$ in round, per-group counts
    for candidate $i$ for samples taken in round, corresponding distance estimates
    using the samples taken in round, respectively} \\ \hline
    $\matchingset, \activeset$ & Set of matching \viztypes{} (see \Cref{def:matching})
    and non-pruned \viztypes, respectively, during a run of \hsim \\ \hline
    \red{$\Ntotali, \Ntotal$}, \blue{$m$, $f(\cdot;\Ntotal,\Ntotali,m)$} & \red{Number of datapoints corresponding to candidate $i$, total number of datapoints}, \blue{samples taken during stage 1, hypergeometric pdf} \\ \hline
    \end{tabular}
    \vspace{-10pt}
  \caption{Summary of notation.}
  \label{tab:notation}
  \vspace{-10pt}

\end{table*}

In this section, we formalize the problem of identifying \viztypes whose
distributions match a reference.  

\subsection{Generation of Histograms}
We start with a concrete example
of the typical database query an analyst might use to generate a \viztype.
Returning to our example from \Cref{sec:intro},
suppose an analyst is interested in studying how 
population proportions vary across income brackets for various countries around the world.
Suppose she wishes to find
countries with populations distributed across different income brackets
most similarly to a specific country, such as Greece.
Consider the following SQL query, where \texttt{\$COUNTRY} is a variable:
\vspace{-3pt}
\begin{quote}
\small
SELECT \texttt{income\_bracket}, \COUNT(*) FROM \texttt{census} \\
WHERE \texttt{country=\$COUNTRY} \\
GROUP BY \texttt{income\_bracket}
\end{quote}
\vspace{-3pt}
This query returns a list of 7 (income bracket, count) pairs to the analyst for
a specific country.  The analyst may then choose to visualize the results by
plotting the counts versus 
different income brackets in a \viztype, i.e.,
a plot similar to the right side of \Cref{fig:interface} (for Italy).  
Currently, the
analyst may examine hundreds of similar \viztypes, one for each country,
comparing it to the one for Greece, to manually identify ones that are similar.

In contrast, the goal of \fm is to perform this search automatically \blue{and} efficiently.
Conceptually, \fm will iterate over all possible values of country, 
generate the corresponding \viztypes,
and evaluate the similarity of its distribution (based on some notion of similarity
described subsequently) to the corresponding visualization for Greece.
In actuality, \fm will perform this search all at once, 
quickly pruning countries that are either clearly close or far from the target.

\topic{Candidate Visualizations} 
Formally, we consider visualizations as being 
generated as a result 
of \textbf{\viztype-generating queries}:

\vspace{-5pt}
\begin{definition}
 \label{def:vizquery}
 A {\em \viztype-generating query} is a SQL query of the following type:
 \vspace{-5pt}
\begin{quote}
\normalfont
\small
 SELECT $X$, \COUNT(*) FROM $T$\\
 WHERE $Z=z_i$ GROUP BY $X$
\end{quote}
\vspace{-5pt}
The table $T$ and attributes $X$ and $Z$
form the query's {\em template}.
\end{definition}
\vspace{-5pt}

For each concrete value $z_i$ of attribute $Z$ specified in the query, 
the results of the query---i.e., the
grouped counts---can be represented in the 
form of a vector $(r_1,r_2,\ldots,r_n)$, where $n =
|V_X|$, the cardinality of the value set of attribute $X$. 
This $n$-tuple can then be
used to plot a \viztype visualization---in this paper, when 
we refer to a \viztype or a visualization, we will be typically referring to such an $n$-tuple. 
For a given \textit{grouping attribute} $X$
and a \textit{candidate attribute} $Z$, we
refer to the set of all visualizations generated by letting $Z$ vary over its
value set as the set of {\bf \em candidate visualizations}.
We refer to each distinct value in the grouping attribute $X$'s
value set as a \textit{group}.
In our example, $X$ corresponds to \texttt{income\_bracket}
and $Z$ corresponds to \texttt{country}.

For ease of exposition, we focus on candidate 
visualizations generated from queries
according to \Cref{def:vizquery}, 
having single \red{categorical} attributes for $X$ and $Z$. Our methods
are more general and extend naturally 
to handle 
{\em (i) predicates:} additional predicates on other attributes,
{\em (ii) multiple and complex $X$s:}
additional grouping (i.e., $X$) attributes,
\red{groups derived from binning real-values
(as opposed to categorical $X$), along with groups derived from binning 
multiple categorical $X$ attribute values together (e.g., quarters instead of individual months)},
and 
{\em (iii) multiple and complex $Z$s:}
additional candidate (i.e., $Z$) attributes,
\red{as well as candidate attribute values derived from binning real values
(as opposed to categorical $Z$)}.
\papertext{We discuss these extensions in our technical report~\cite{techreport}.}
The flexibility in specifying \viztype-generating
queries---exponential in the number of attributes---makes it
impossible for us to precompute the results of all such queries.

\topic{Visualization Terminology}
Our methods are agnostic to the particular method used to present visualizations.
That is, analysts may choose to present the results generated from queries
of the form in \Cref{def:vizquery} via line plots, heat maps,
choropleths, and other visualization types, as any of
these may be specified by an ordered tuple of real values and are thus
permitted under our notion of a ``candidate visualization''.
We focus on bar charts of frequency counts and \viztypes---these
naturally capture aggregations over the categorical or binned quantitative 
grouping attribute $X$ respectively.
Although a bar graph plot of frequency counts over a categorical grouping attribute
is not technically a histogram, which implies that the grouping attribute is continuous,
we loosely use the term ``histogram'' to refer to both cases in a unified way.

\topic{Visual Target Specification}
Given our specification of candidate visualizations, 
a {\bf \em visual target} is an
$n$-tuple, denoted by $\vQ$ with entries $Q_1, Q_2, \ldots,
Q_n$, that we need to match the candidates with.  
Returning to our flight
delays example, $\vQ$ would refer to the visualization corresponding to
Greece, with $Q_1$ being the count of individuals in the first income bracket,
$Q_2$ the
count of individuals in the second income bracket,
and so on.  

\topic{Samples}
To estimate these candidate visualizations, we need to take {\em samples}.
In particular, for a given candidate $i$ for some attribute $Z$, a sample
corresponds to a single tuple $t$ with attribute value $Z=z_i$.
The attribute value $X=x_j$ of $t$ increments the $j$th entry
of the estimate $\vri$ for the candidate \viztype.

\topic{Candidate Similarity} 
Given a set of candidate visualizations with estimated vector
representations $\{\vri\}$
such that the $i$th candidate is generated by selecting on $Z=z_i$, our
problem hinges on finding the candidate whose distribution is
most ``similar'' to the visual target $\vQ$
specified by the analyst. For quantifying visual similarity, we do
not care about the absolute counts $r_1,r_2,\ldots,r_\mvx$, and
instead prefer to determine whether $\vri$ and $\vQ$ are close
in a {\em distributional} sense. Using hats to denote normalized
variants of $\vri$ and $\vQ$, write
\[ \vhri = \frac{\vri}{\mathbf{1}^T\vri} \qquad \vhQ = \frac{\vQ}{\mathbf{1}^T\vQ} \]
With this notational convenience, we make our notion of similarity explicit
by defining candidate distance as follows:
\vspace{-5pt}
\begin{definition}
 \label{def:cand_distance}
 For candidate $\vri$ and visual predicate $\vQ$, the \textbf{distance}
 $d(\vri,\vQ)$ between $\vri$ and $\vQ$ is defined as follows:
 \[ d(\vri, \vQ) = \norm{\vhri - \vhQ}_1 = \norm{\frac{\vri}{\mathbf{1}^T\vri} - \frac{\vQ}{\mathbf{1}^T\vQ}}_1 \]
\end{definition}
\vspace{-5pt}
That is, after normalizing the candidate and target vectors so that their
respective components sum to 1 (and therefore correspond to distributions), we
take the $\ell_1$ distance between the two vectors.
When the target $\vQ$ is understood from context, we denote the distance
between candidate $\vri$ and $\vQ$ by $\tau_i = d(\vri, \vQ)$.

\techreport{
\begin{figure*}[t]
\centering
\includegraphics[width=.3\textwidth]{\figs/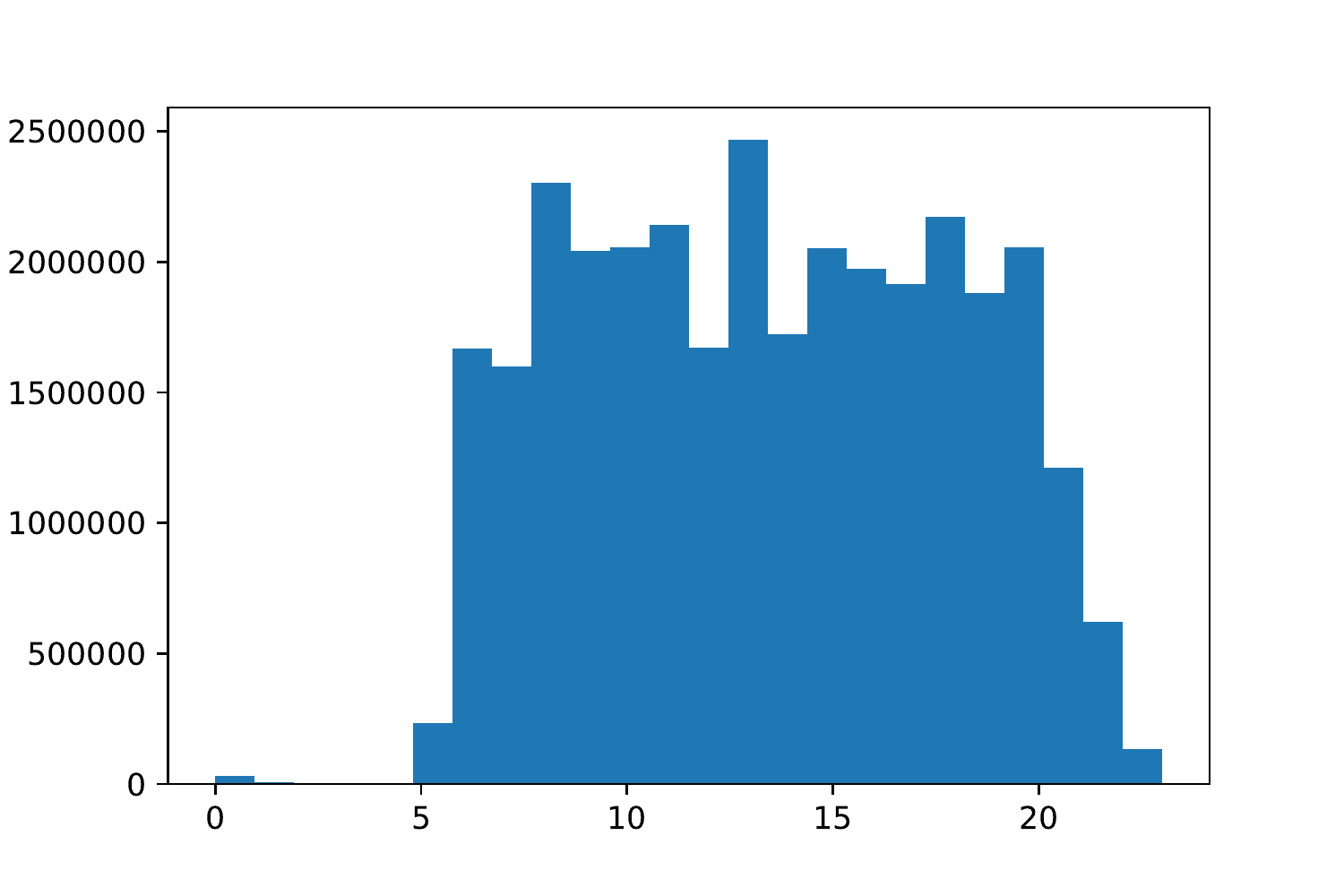}
\includegraphics[width=.3\textwidth]{\figs/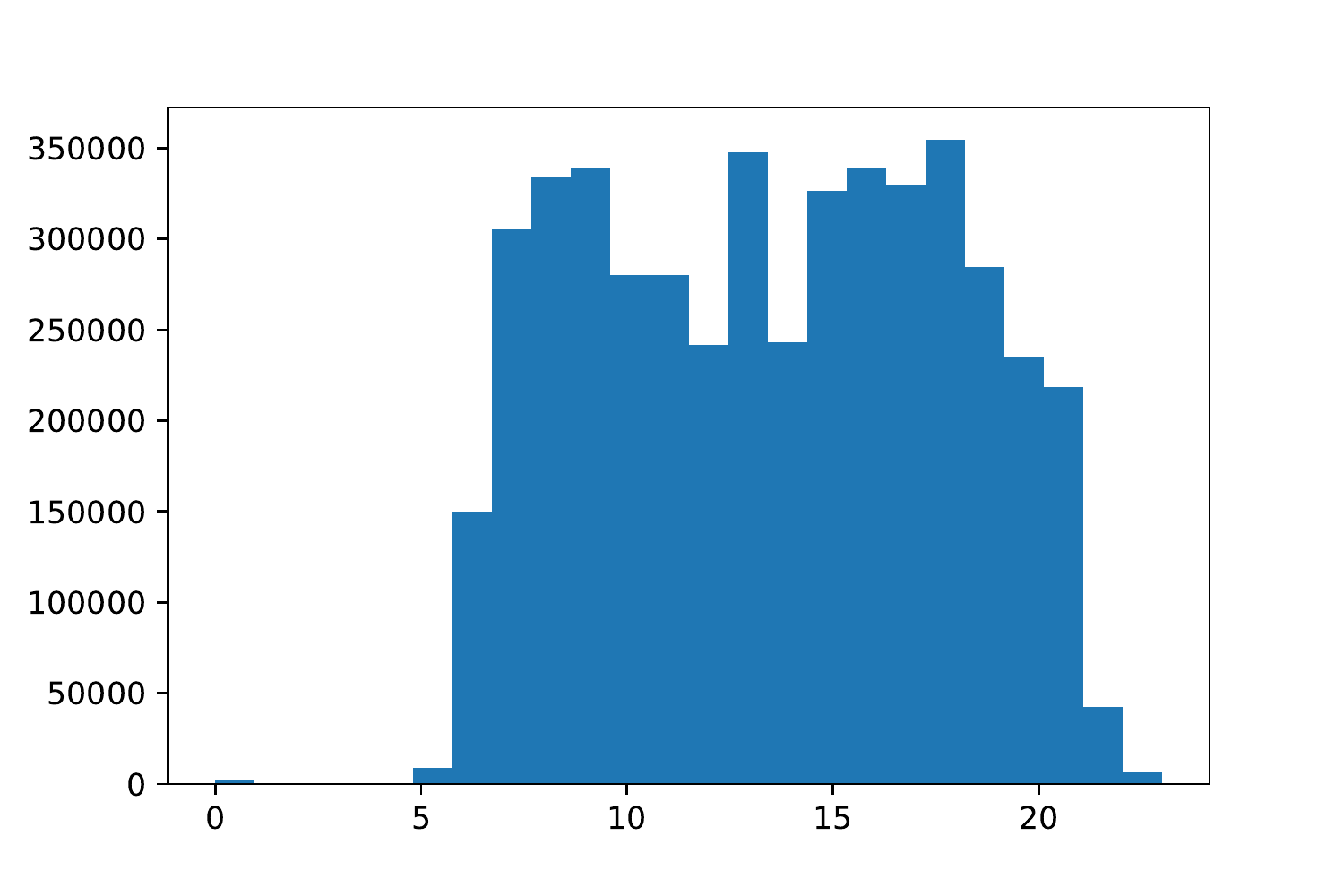}
\includegraphics[width=.3\textwidth]{\figs/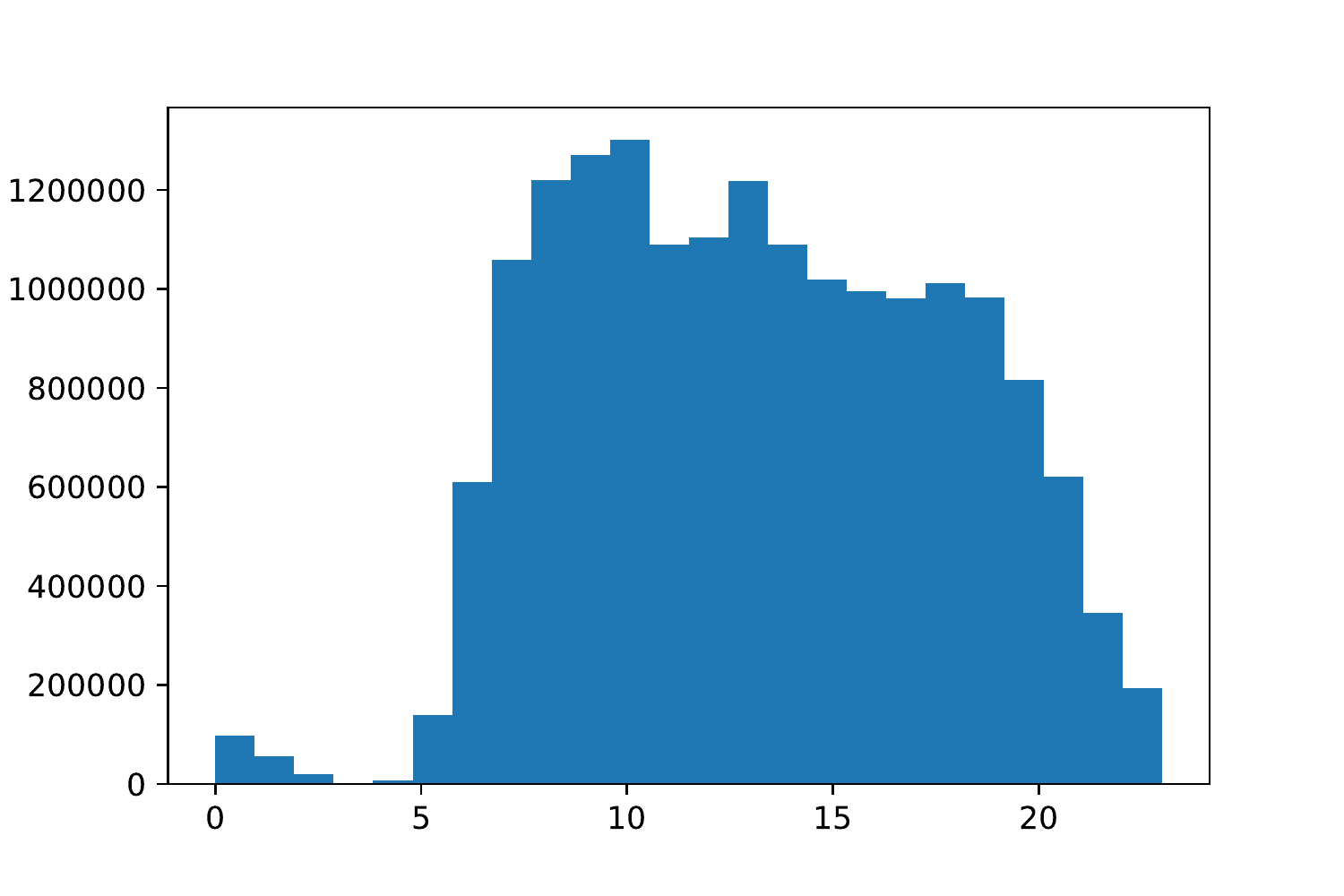}
\vspace{-10pt}
\caption{\red{The target (departure hour histogram for ORD), second closest in normalized $\ell_1$ (DAL) , second closest in normalized $\ell_2$ (PHX)}}
\label{fig:l1l2comp}
\vspace{-10pt}
\end{figure*}

\begin{figure}[t]
\centering
\vspace{-8pt}
\includegraphics[width=.3\textwidth]{\figs/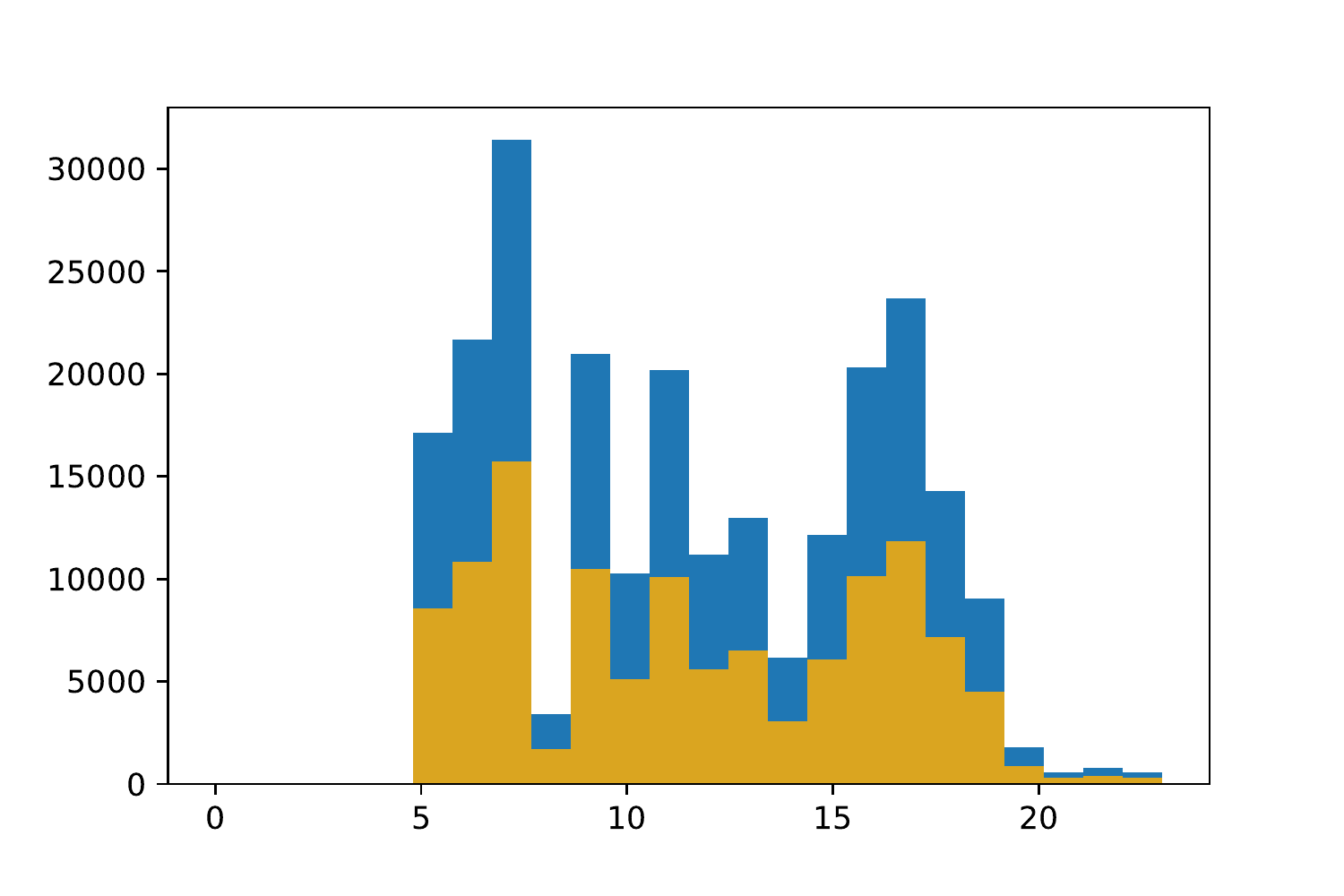}
\vspace{-10pt}
\caption{\red{The goldenrod histogram is identical to the blue one post-normalization,
but appears very far visually pre-normalization.}}
\label{fig:normalize}
\end{figure}
}

\topic{The Need for Normalization}
A natural question that readers may have is why we chose to normalize each
vector prior to taking the distance between them.  We do this because
 the goal of \fm is to find visualizations that have
similar distributions, as opposed to similar actual values. 
Returning to our example, if we consider the population
distribution of Greece
across different income brackets, and compare it to that of other
countries, without normalization, we will end up returning other countries with
similar population counts in each bin---e.g., other countries with
similar overall populations---as opposed to those that have similar shape or
distribution. \techreport{\red{To see an illustration of this, consider
\Cref{fig:normalize}. The overlaid histogram in goldenrod is identical to
the blue one, but we are unable to capture this without normalization.}

\topic{\red{Choice of Metric Post-Normalization}}}
A similar metric, using $\ell_2$ distance between
normalized vectors (as opposed to $\ell_1$), has been studied in prior
work~\cite{vartak2015s,SPS} and even validated in a user study
in~\cite{vartak2015s}. However, as observed
in~\cite{BatuTestingDistributionsAreClose}, the $\ell_2$ distance between
distributions has the drawback that it could be small even for distributions
with disjoint support.
The $\ell_1$ distance metric over discrete probability
distributions has a direct correspondence with the traditional statistical
distance metric known as {\em total variation distance}~\cite{gibbs2002choosing}
and does not suffer
from this drawback\papertext{, so we prefer it in this work.}
\techreport{.

\red{Additionally, we sometimes observe that $\ell_2$ heavily penalizes
candidates with \blue{a small number of vector entries with large deviations
from each other}, even when they are arguably
closer visually than those candidates closest in $\ell_2$.
Consider \Cref{fig:l1l2comp}, which depicts histograms generated by one of
the queries on a \flights dataset we used in our experiments,
corresponding to a histogram of departure time. The target is the Chicago ORD airport,
and we are depicting the first non-ORD top-k histogram for both $\ell_1$ and $\ell_2$ (i.e., the 2nd ranked
histogram for both metrics), among all airports.
As one can see in the figure, 
the middle histogram is arguably ``visually closer'' to the ORD histogram on the left, but is not considered
so by $\ell_2$ due to the mismatch at about the 6th hour.}}

\techreport{\red{KL-divergence is another possibility as a distance metric, but it
has the drawback that it will be infinite for any candidate that places 0 mass in a place where the
target places nonzero mass, making it difficult to compare these (note that this follows directly from the definition:
$KL(p\Vert q) = -\sum_i p_i\log \frac{q_i}{p_i}$).}}

\subsection{Guarantees and Problem Statement}
\newcommand{\separation}{\textsc{Separation}\xspace}
\newcommand{\reconstruction}{\textsc{Reconstruction}\xspace}
\newcommand{\probcorrect}{\textsc{Top-K-Similar}\xspace}

Since \fm takes samples to estimate the candidate
\viztype visualizations, and therefore may return incorrect results, 
we need to enforce probabilistic guarantees on the correctness
of the returned results.

First, we introduce some notation:
we use $\vri$ to denote the {\em estimate} of the candidate
visualization, while $\vrsi$ (with normalized version $\vhrsi$) is the {\em true}
candidate visualization on 
the entire dataset. 
Our formulation also relies on constants $\veps$, $\delta$, \red{and $\sigma$}, which we
assume either built into the system or provided by the analyst.
\red{We further use $\Ntotal$ and \blue{$\Ntotali$} to denote the total
number of datapoints and number of datapoints corresponding
to candidate \blue{$i$}, respectively.}
\vspace{-2pt}
\begin{guarantee} (\separation)\xspace
\label{guarantee:separation}
Any \blue{approximate} \viztype \blue{$\vri$ with selectivity
$\frac{\Ntotali}{\Ntotal}\geq\minsel$}
that is in the true top-$k$ closest (w.r.t.
\Cref{def:cand_distance}) but {\em not} part of the output
will be less than $\veps$ closer to the target
than the furthest \viztype that {\em is} part of the output.
That is, if the algorithm outputs \viztypes $\vr_{j_1}, \vr_{j_2}, \ldots, \vr_{j_k}$,
then, \blue{for all $i$}, $\max_{1\leq l \leq k}\left\{d(\vrs_{j_l}, \vQ)\right\} - d(\vrsi, \vQ) < \veps, $
\red{or $\frac{\Ntotali}{\Ntotal} < \minsel$.}
\end{guarantee}
\vspace{-3pt}
\techreport{\red{Note that we use ``selectivity'' as a number and not as a property,
matching typical usage in database systems literature~\cite{selinger1979access,kester2017access}.
As such, candidates with lower selectivity appear less frequently in the data than
candidates with higher selectivity.}\agpres{Why are you not calling it frequency?
Selectivity is always confusing.}\resolvedres{As we discussed, the Serfling bound does
not do will with a frequency or a minsup if the data gets big enough, so selectivity
is the most appropriate to avoid eyebrow-raising of changing minimum frequency thresholds.}}
\vspace{-5pt}
\begin{guarantee} (\reconstruction)\xspace
\label{guarantee:reconstruction}
Each approximate \viztype $\vri$ 
output as one of the top-$k$ satisfies
$d(\vri, \vrsi) < \veps$.
\end{guarantee}
\vspace{-5pt}

The first guarantee says that any ordering mistakes are relatively innocuous:
for any two \viztypes $\vri$ and $\vrj$, if the algorithm
outputs $\vrj$ but not $\vri$, when it should have been the other way around,
then either $\left|d(\vrsi, \vQ) - d(\vrsj, \vQ)\right| < \veps$, \red{or $\frac{\Ntotali}{\Ntotal} < \minsel$.
The intuition behind the minimum selectivity parameter, $\minsel$, is that certain candidates
may not appear frequently enough within the data to get a reliable reconstruction of the
true underlying distribution responsible for generating the original data,
and thus may not be suitable for downstream decision-making. 
For example, in our income example, a country with a population of 100 may have a histogram
similar to the visual target 
but this would not be statistically significant.}
Overall, our guarantee states that we still return a visualization that is quite close to $\vQ$,
\red{and we can be confident that anything dramatically closer
has relatively few total datapoints available within the data (i.e., $\Ntotali$ is small).}

The second guarantee says that the \viztypes output are
not too dissimilar from the corresponding true distributions that would result from a
complete scan of the data. As a result, they form an adequate and accurate proxy from which
insights may be derived. 
With these definitions in place, we now formally state our core problem:
\vspace{-5pt}
\begin{problem} (\probcorrect).
\label{prob:topk-correct}
Given a visual target $\vQ$, a \viztype-generating
query template, $k$, $\veps$, $\delta$, \red{and $\sigma$},
display $k$ candidate attribute values $\{z_i\} \subseteq
\vz$ (and accompanying visualizations $\{\vri\}$)
as quickly as possible, such that the output
satisfies \guarantees with probability
greater than $1-\delta$.
\end{problem}
\vspace{-5pt}

\densesection{The {\large \hsim} Algorithm}
\label{sec:algorithm}

\noindent In this section, we discuss how to conceptually solve
\Cref{prob:topk-correct}.
We outline an algorithm, named \hsim,
which allows us to determine confidence levels for whether our separation and reconstruction
guarantees hold. \red{\blue{We rigorously prove in this section that when our algorithm terminates, it
gives correct results with probability greater than $1-\delta$ {\em regardless} of the data given as input.}}
Many systems-level details \red{and other heuristics used to make \hsim
perform \blue{particularly well in practice}} will be presented in 
\Cref{sec:arch}. \Cref{tab:notation}
 provides a description of the notation used. \papertext{All proofs
 appearing in this section can be found in our technical report~\cite{techreport}.}

\subsection{Algorithm Outline}
\label{subsec:fastmatch}

\newcommand{\pluseq}{\mathrel{+}=}
\begin{algorithm}[t]
 \caption{The \hsim algorithm}\label{alg:basic}
\SetKwInOut{Input}{Input}
\SetKwInOut{Output}{Output}
\SetKwRepeat{Do}{do}{while}
{\scriptsize

\Input{Columns $Z, X$, visual target $\vQ$, parameters $k, \veps, \delta, \sigma$}
\Output{Estimates $\matchingset$ of the top-$k$ closest candidates to $\vQ$, histograms $\{\vri\}$}
\ \\
{\bf Initialization.}\ \\
$\takenfori, \blue{\takenforip} \gets 0$, \blue{$\vri, \vrip \gets \vzero$} for $1\leq i\leq \mvz$\;
\ \\
{\bf stage 1:} $\deltaupper \gets \frac{\delta}{\blue{3}}$\;
Repeat \blue{$m$ times: uniformly randomly sample some tuple without replacement}\;
\label{hsim:line:sample-stage1}
Update $\{\takenfori\}$, $\{\vri\}$, \blue{$\{\taui\}$} based on the new samples\;
\label{hsim:line:update-stage1}
\label{hsim:line:update-delta-start-stage1}
   \blue{$\Delta \gets \{\delta_i\}$ where $\delta_i = \sum_{j=0}^{\takenfori} f(j; N,\lceil\minsel N\rceil, m)$ for $1\leq i\leq\mvz$}\;
\label{hsim:line:update-delta-end-stage1}
Perform a Holm-Bonferroni statistical test with P-values in $\Delta$; that is:\\
\label{hsim:line:holmbon-test-stage1}
\blue{$\activeset \gets \left\{i : \delta_i\leq\frac{\delta}{\mvz-i+1}\text{ and for all $j<i$, } \delta_j\leq\frac{\delta}{\mvz-j+1}\right\}$}\;
\ \\
{\bf stage 2:} $\deltaupper \gets \frac{\delta}{\blue{3}}$\;
\Do{\ \blue{$\max(\Delta)>\deltaupper$}\label{hsim:line:multitest-stage2}}{
  $\deltaupper \gets \frac{1}{2}\delta^{upper}$\;
  \blue{$\takenfori \pluseq \takenforip$, $\quad \vri \pluseq \vrip$, $\quad \taui \gets d(\vri, \vQ) \quad$ for $i\in A$}\;
  \label{hsim:line:accum-stage2-start}
  \blue{$\takenforip \gets 0$, $\quad \vrip \gets \vzero \quad$ for $i\in A$}\;
  \label{hsim:line:accum-stage2-end}
  \blue{$M \gets \{i\in A : \taui\text{ among $k$ smallest}\}$}\;
  \label{hsim:line:matchingset}
  $s \gets \frac{1}{2}(\max_{i\in\matchingset} \tau_i  + \min_{j\in\activeset\setminus\matchingset} \tau_j)$\;
  \label{hsim:line:splitchoice}
  Repeat: take uniform random samples from any $i\in \activeset$\;
  \label{hsim:line:sample-stage2}
  \blue{Update $\{\takenforip\}$, $\{\vrip\}$, and $\{\tauip\}$ based on the new samples}\;
  \label{hsim:line:update-stage2}
  \label{hsim:line:multitest-stage2-start}
  \blue{$\veps_i \gets s+\frac{\veps}{2}-\tauip$ for $i \in \matchingset$}\;
  \blue{$\veps_j \gets \taujp-(s-\frac{\veps}{2})$ if $s-\frac{\veps}{2}\geq 0$ else $\infty$ for $j \in \activeset\setminus\matchingset$}\;
  \blue{$\Delta \gets \{\delta_i\}$ where $\delta_i \geq \Parg{d(\vrip, \vrsi) > \veps_i}$ for $i\in\activeset$}\;
  \label{hsim:line:multitest-stage2-end}
}

\ \\
\blue{{\bf stage 3:} Sample until $n_i \geq \frac{2}{\veps^2}\left(\mvx\log{2} + \log{\frac{3k}{\delta}}\right)$, for all $i\in\matchingset$}\;
\label{hsim:line:sample-stage3}
\blue{Update $\{\vri\}$ based on the new samples}\;
\Return{$\matchingset$, and $\{\vri \blue{: i\in\matchingset}\}$}\;
}
\end{algorithm}

\begin{figure}[t]
\vspace{-10pt}
 \begin{center}
  \includegraphics[width=.4\textwidth]{\figs/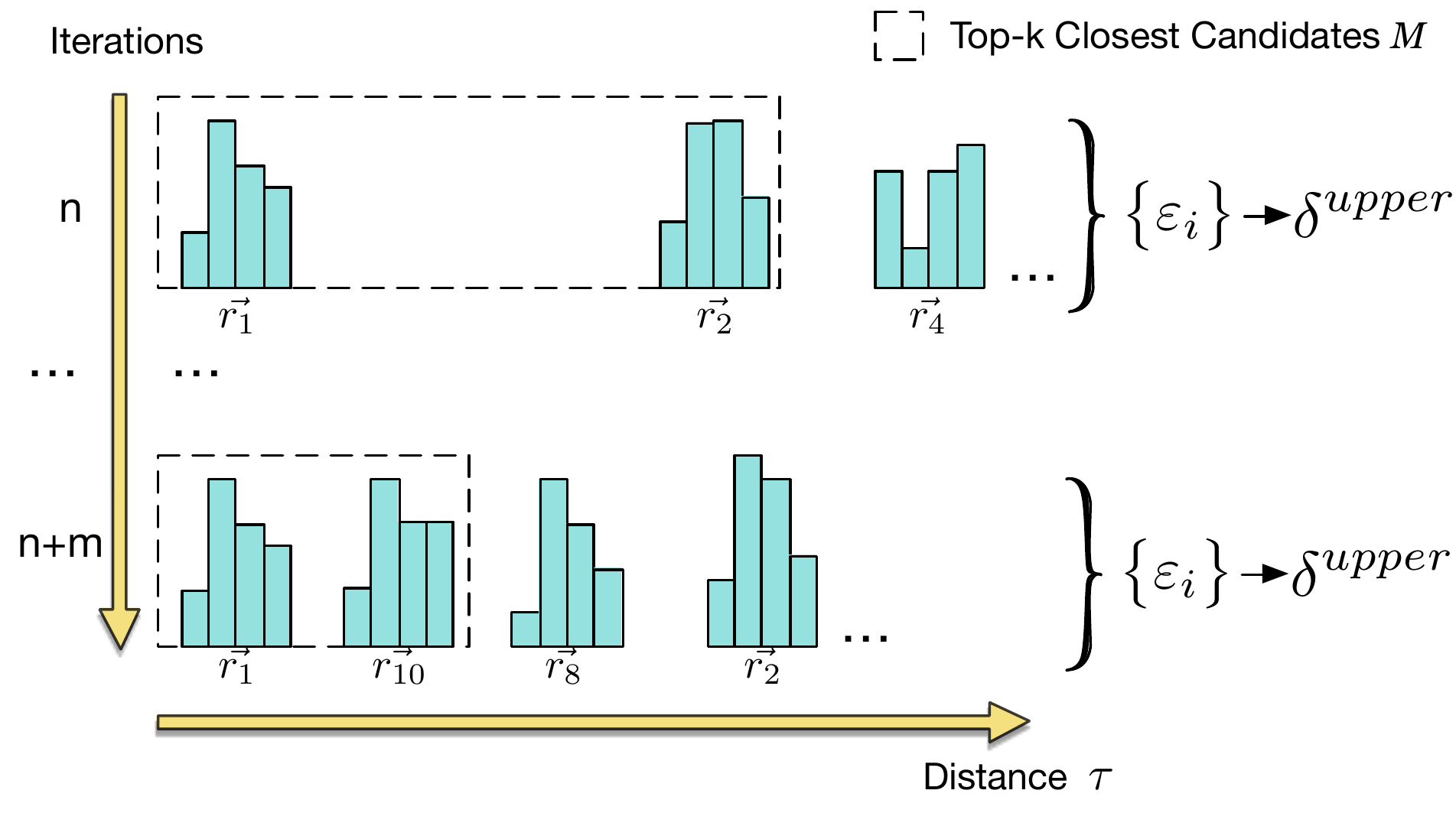}
  \vspace{-10pt}
  \caption{Illustration of \hsim.}
  \label{fig:fmtopk}
 \end{center}
 \vspace{-1em}
\end{figure}

\hsim operates by sampling tuples. Each of these tuples contributes to one or
more candidate \viztypes, using which \hsim constructs \viztypes
$\{\vhri\}$. After taking enough samples corresponding to each candidate, it
will eventually be likely that $d(\vri,\vrsi)$ is ``small'', and that
$|d(\vri, \vQ) - d(\vrsi, \vQ)|$ is likewise ``small'', for each $i$. More
precisely, the set of candidates will likely be in a state such that
\guarantees are both satisfied simultaneously.

\topic{\blue{Stages Overview}}
\red{\hsim separates its sampling into \blue{three} stages,
each with an error probability of at most $\frac{\delta}{3}$,
giving an overall error probability of at most $\delta$:}
\begin{denselist}
\item \red{Stage 1 \blue{[Prune Rare Candidates]}: Sample datapoints uniformly at random \blue{without replacement}, so that each candidate
is sampled a number of times \blue{roughly} proportional to the number of datapoints
corresponding to that candidate. Identify rare candidates that likely satisfy
$\frac{\Ntotali}{\Ntotal} < \minsel$, and prune these ones.}
\agpres{A skeptic will say: if you know how to bias this perfectly, you already know $\Ntotali$.
Instead maybe say: sample datapoints uniformly at random? (Or something equivalent.)}
\resolvedres{Tried to make this clearer above.}
\item \red{Stage 2 \blue{[Identify Top-$k$]}:
\blue{Take samples from the remaining candidates}
until the top-$k$ have been identified reliably.}
\item \blue{Stage 3 \blue{[Reconstruct Top-$k$]}: Sample from the estimated top-$k$
until they have been reconstructed reliably.}
\end{denselist}
\red{This separation is important for performance: the pruning step
(stage 1) often {\em dramatically reduces} the number of candidates
that need to be considered in \blue{stages 2 and 3}.}

\red{\blue{The first two} stages of \hsim \blue{factor into phases that are pure I/O and phases that involve
one or more statistical tests.}
\blue{The I/O phases sample tuples} (lines~\ref{hsim:line:sample-stage1} and~\ref{hsim:line:sample-stage2}
in Algorithm~\ref{alg:basic})---we will
describe how in Section~\ref{sec:arch}; our algorithm's correctness is independent of
how this happens, \blue{provided that the samples are uniform.}}

\topic{\blue{Stage 1: Pruning Rare Candidates (\Cref{subsec:stage1})}}
\red{During stage 1, the I/O phase (line~\ref{hsim:line:sample-stage1})
\blue{takes $m$ samples, for some $m$ fixed ahead of time.}
This is followed by updating, for each candidate $i$,
the number of samples $\takenfori$ observed so far (line~\ref{hsim:line:update-stage1}), 
and using the \blue{P-values} $\{\delta_i\}$ \blue{of a test for underrepresentation
to determine} whether \blue{each candidate $i$
is rare, i.e., has $\frac{\Ntotali}{\Ntotal} < \minsel$}
(lines~\ref{hsim:line:update-delta-start-stage1}--\ref{hsim:line:holmbon-test-stage1}).}

\topic{\blue{Stage 2: Identifying Top-$k$ (\Cref{subsec:stage2})}}
\red{For stage 2, we focus on a smaller set of candidates;
namely, those that we did not find to be \blue{rare} 
(denoted by $\activeset$).
\blue{Stage 2 is divided into {\em rounds}.
Each round attempts to use existing samples to estimate which
candidates are top-$k$ and which are non top-$k$, and then
draws new samples, testing how unlikely it is to observe the
new samples in the event that its guess of the top-$k$ is wrong.
If this event is unlikely enough, then it has
recovered the correct top-$k$, with high probability.}}

\red{\blue{At the start of each round, \hsim accumulates any samples taken
during the previous round
(lines~\ref{hsim:line:accum-stage2-start}--\ref{hsim:line:accum-stage2-end}).
It then determines the current top-$k$ candidates and
a separation point $s$ between top-$k$ and non top-$k$
(lines~\ref{hsim:line:matchingset}--\ref{hsim:line:splitchoice}),
as this separation point determines a set of hypotheses to test.
Then, it begins an I/O phase and takes samples (\cref{hsim:line:sample-stage2}).
The samples taken each round
are used to generate the number of samples taken per candidate,
$\{\takenforip\}$, the estimates
$\{\vrip\}$, and the distance estimates $\{\tauip\}$ (\cref{hsim:line:update-stage2}).
These statistics are computed from fresh samples
each round (i.e., they do not reuse samples across rounds) so that they may
be used in a statistical test
(lines~\ref{hsim:line:multitest-stage2-start}--\ref{hsim:line:multitest-stage2-end}),
discussed in \Cref{subsec:stage2}.
After computing the P-values for each null hypothesis to test
(\cref{hsim:line:multitest-stage2-end}), \hsim
determines whether it can reject all the hypotheses with type 1 error
(i.e., probability of mistakenly rejecting a true null hypothesis) bounded
by $\deltaupper$ and break from the loop (\cref{hsim:line:multitest-stage2}).
If not, it repeats with new samples and a smaller $\deltaupper$ (where the $\{\deltaupper\}$
are chosen so that the probability of error across {\em all} rounds is at most $\frac{\delta}{3}$).}}

\topic{\blue{Stage 3: Reconstructing Top-$k$ (Section~\ref{subsec:overall-proof})}}
\blue{Finally, stage 3 ensures that the identified top-$k$, $M$, all satisfy
$d(\vri, \vrsi) \leq \veps$ for $i\in\matchingset$ (so that \grec holds),
with high probability.}

\smallskip

\noindent \Cref{fig:fmtopk} illustrates \hsim \red{stage 2} running on a toy example in which
we compute the top-2 closest \viztypes to a target. At round $n$, it estimates $\vr_1$
and $\vr_2$ as the top-2 closest, which it refines by the time it reaches round $n+m$.
As the rounds increase, \smackoutres{and \hsim takes more samples,
$\deltaupper$ decreases}{\hsim takes more samples \blue{to get better estimates
of the distances $\{\taui\}$ and thereby improve the chances of termination
when it performs its multiple hypothesis test in stage 2.}}

\topic{Choosing where to sample and how many samples to take}
\blue{The estimates $\matchingset$ and $\{\taui\}$ allow us to}
determine which candidates are
``important'' to sample from in order to allow termination with fewer samples;
we return to this in \Cref{sec:arch}.
Our \hsim algorithm is agnostic to the sampling approach.

\topic{Outline}
\red{We first discuss the Holm-Bonferroni method for testing multiple statistical hypotheses
simultaneously in \Cref{subsec:holmbon}, \blue{since stage 1 of} \hsim \blue{uses} it as a subroutine,
\blue{and since the simultaneous test in stage 2 is based on similar ideas.}
In \Cref{subsec:stage1}, we discuss stage 1 of \hsim, and prove that upon termination,
all candidates $i$ flagged for pruning satisfy $\frac{\Ntotali}{\Ntotal} < \minsel$ with probability
greater than $\frac{\delta}{\blue{3}}$. Next, in \Cref{subsec:stage2}, we discuss stage 2 of \hsim, and prove
that upon termination, we have the guarantee that any non-pruned candidate mistakenly
classified as top-$k$ is no more than $\veps$ further from the target than the furthest
true non-pruned top-$k$ candidate \blue{(with high probability)}. The proof of correctness for stage 2
is the most involved and is divided as follows:
\begin{denselist}
\item In \Cref{subsec:deviations-implies-guarantees}, we 
\blue{give lemmas that allow us to relate the reconstruction of the candidate
histograms from estimates $\{\vrip\}$ to the separation guarantee via multiple hypothesis testing;}
\item In \Cref{subsec:selecting-deviations}, we describe a method to select
\blue{appropriate hypotheses
to use for testing in the lemmas of \Cref{subsec:deviations-implies-guarantees};}
\item In \Cref{subsec:confidence}, 
we prove a theorem that enables us to use the samples per candidate \viztype
to \blue{determine the P-values associated with the hypotheses.}
\end{denselist}}
\noindent In \Cref{subsec:overall-proof},
we \blue{discuss stage 3 and} conclude with an overall proof of correctness.

\subsection{Controlling Family-wise Error}
\label{subsec:holmbon}
\red{In \blue{the first two stages} of \hsim, the algorithm needs to perform multiple statistical tests
simultaneously~\cite{casella2002statistical}.
In stage 1, \hsim tests null hypotheses of the form ``candidate $i$ is high-selectivity''
versus alternatives like ``candidate $i$ is not high-selectivity''. In this case, ``rejecting the null hypothesis
at level $\deltaupper$'' roughly means that the probability that candidate $i$
is high-selectivity is at most $\deltaupper$.
Likewise, during stage 2, \hsim tests null hypotheses of the form ``candidate $i$'s
\blue{true distance from $\vQ$, $\tau_i^*$, lies above (or below) some fixed value $s$.''
If the algorithm correctly rejects every null hypothesis while controlling the family-wise
error~\cite{lehmann2006testing} at level $\deltaupper$, then it
has correctly determined which side of $s$ every
$\tau_i^*$ lies, a fact that we use to get the separation guarantee.}}

\red{\blue{Since stages 1 and 2} test multiple hypotheses at the same time,
\hsim needs to control the family-wise type 1 error (false positive) rate of
these tests simultaneously. That is, if the family-wise type 1 error is controlled
at level $\deltaupper$, then the probability that one or more rejecting tests in the
family should not have rejected is less than $\deltaupper$ --- during stage 1, this intuitively means
that the probability one or more high-selectivity candidates were deemed to be low-selectivity is
at most $\deltaupper$, and during stage 2, this roughly means that the probability
\blue{of selecting some candidate as top-$k$ when it is non top-$k$ (or vice-versa)
is at most $\deltaupper$.}}

\red{The reader \blue{may be}
familiar with the Bonferroni correction, which
\blue{enforces a family-wise error rate of $\deltaupper$ by requiring a significance
level $\frac{\deltaupper}{\mvz}$ for each test in a family with $\mvz$ tests in total.}
\blue{We instead use} the Holm-Bonferroni method~\cite{holm1979simple}, \blue{which}
is uniformly more powerful than the Bonferroni correction, meaning that it needs
fewer samples to make the same guarantee. \techreport{Like its simpler counterpart, it is
correct regardless of whether the family of tests has any underlying dependency structure.} In brief,
a level $\deltaupper$ test over a family of size $\mvz$ works by first
sorting the P-values $\{\delta_i\}$
of the individual tests in increasing order,
and then finding the minimal index $j$ (starting from 1) where $\delta_j > \frac{\deltaupper}{\mvz-j+1}$
(if this does not exist, then set $j=\mvz$).
The tests with smaller indices reject their respective null hypotheses at level $\deltaupper$, and
the remaining ones do not reject.}

\subsection{Stage 1: Pruning Rare Candidates}\label{subsec:stage1}
\techreport{\blue{One way to remove {\em rare} (i.e. low-selectivity) candidates from processing is to
use an index to look up how many tuples correspond to each candidate.
While this will work well for some queries, it unfortunately does not
work in general, as candidates generated from queries of the form in
\Cref{def:vizquery} could have arbitrary predicates attached, which cannot
all be indexed ahead-of-time. Thus, we turn to sampling.}}

\blue{To prune rare candidates, we need some way to determine whether
each candidate $i$ satisfies $\frac{\Ntotali}{\Ntotal} < \minsel$ with
high probability. To do so, we make the simple observation that, after drawing
$m$ tuples without replacement uniformly at random, the number of tuples corresponding
to candidate $i$ follows a hypergeometric distribution~\cite{johnson2005univariate}.
The number of samples to take,
$m$, is a parameter; we observe in our experiments that $m=5\cdot10^5$ is an appropriate
choice.\footnote{\scriptsize \blue{Our results are not sensitive to the choice of $m$,
provided $m$ is not too small (so that the algorithm fails to
prune anything) or too big (i.e., a nontrivial fraction of the data).}}
That is, if candidate $i$ has
$\Ntotali$ total corresponding tuples in a dataset of size $\Ntotal$, then the
number of tuples $\takenfori$ for candidate $i$ in a uniform sample without replacement
of size $m$ is distributed according to $\takenfori \sim \HypGeo(\Ntotal, \Ntotali, m)$.
As such, we can make use of a well-known test for underrepresentation~\cite{lehmann2006testing}
to accurately
detect when candidate $i$ has $\frac{\Ntotali}{\Ntotal} < \minsel$. The null hypothesis
is that candidate $i$ is not underrepresented (i.e., has $\Ntotali \geq \minsel\Ntotal$),
and letting $f(\ \cdot\ ;\Ntotal, \lceil\minsel\Ntotal\rceil, m)$
denote the hypergeometric pdf in this case, the P-value for the test is given by
\papertext{$\sum_{j=0}^{\takenfori} f(j; \Ntotal, \lceil\minsel\Ntotal\rceil, m)$,}
\techreport{\[\sum_{j=0}^{\takenfori} f(j; \Ntotal, \lceil\minsel\Ntotal\rceil, m)\]}
where $\takenfori$ is the number of observed tuples for candidate $i$ in the sample of size $m$.
Roughly speaking, the P-value measures how surprised we are to observe $\takenfori$ or fewer
tuples for candidate $i$ when $\frac{\Ntotali}{\Ntotal} \geq \minsel$ --- the lower the
P-value, the more surprised we are.}

\blue{If we reject the null hypothesis for some candidate $i$ when the P-value is at most $\delta_i$,
we are claiming that candidate $i$ satisfies $\frac{\Ntotali}{\Ntotal}<\minsel$, and the
probability that we are wrong is then at most $\delta_i$. Of course, we need to test {\em every}
candidate for rareness, not just a given candidate, which is why \hsim stage 1 uses a Holm-Bonferroni
procedure to control the {\em family-wise} error at any given threshold. \techreport{We note in passing that
the joint probability of observing $\takenfori$ samples for candidate $i$ across {\em all} candidates
is a multivariate hypergeometric distribution for which we could perform a similar test without
a Holm-Bonferroni procedure, but the CDF of a multivariate hypergeometric is extremely expensive
to compute, and we can afford to sacrifice some statistical power for the sake of computational efficiency
since we only need to ensure that the candidates pruned are actually rare, without necessarily
finding all the rare candidates --- that is, we need high precision, not high recall.
\smacke{Talk about how sublinear algorithms for finding frequent items do not apply.}}}

\red{We now prove a lemma regarding correctness of stage 1.}
\begin{lemma}[Stage 1 Correctness]
 \label{lem:stage1correct}
 \red{After \hsim stage 1 completes, every candidate $i$ removed from $\activeset$
 satisfies $\frac{\Ntotali}{\Ntotal} < \minsel$, with probability greater than $1-\frac{\delta}{\blue{3}}$}
\end{lemma}
\techreport{\begin{proof}
\blue{This follows immediately from the above discussion, in conjunction with the fact that
the P-values generated from each test for underrepresentation are fed into a Holm-Bonferroni
procedure that operates at level $\frac{\delta}{\blue{3}}$, so that the probability of pruning
one or more non-rare candidates is bounded above by $\frac{\delta}{\blue{3}}$.}
\end{proof}}
\papertext{\red{The proof is \blue{a consequence of the correctness of each
individual test for underrepresentation in conjunction with the correctness of the Holm-Bonferroni
procedure for family-wise error~\cite{techreport}.}}}

\subsection{Stage 2: Identifying Top-$K$ Candidates}
\label{subsec:stage2}
\red{\hsim stage 2 attempts to find the top-$k$ closest to the
target out of those remaining \blue{after stage 1. To facilitate discussion,}
we first introduce some definitions.}
\vspace{-4pt}
\begin{definition}(\textsc{Matching Candidates})
\label{def:matching}
A candidate 
is called {\em matching} if
its distance estimate $\taui = d(\vri, \vQ)$
is among the $k$ smallest out of all candidates remaining \blue{after stage 1}.
\end{definition}
\vspace{-3pt}
We denote the \blue{(dynamically changing)} set of candidates that are matching 
during a run of \hsim as $\matchingset$;
we likewise denote the true set of matching candidates
\red{out of the remaining, non-pruned candidates in $\activeset$} as $\matchingset^*$. 
Next, we introduce the notion of {\em \vepsi-deviation}.
\vspace{-8pt}
\begin{definition}(\vepsi-\textsc{deviation})
\label{def:deviance}
\blue{The empirical vector of counts $\vri$ for some} candidate $i$
has \vepsi-deviation if the \blue{corresponding normalized vector}
$\vhri$ \blue{is within} $\veps_i$
\blue{of} the exact distribution $\vhrsi$.
That is, $d(\vri, \vrsi) = \norm{\vhri - \vhrsi}_1 < \veps_i $
\end{definition}
Note that \Cref{def:deviance} overloads the symbol $\veps$ 
to be candidate-specific by appending a subscript. 
In \Cref{subsec:confidence}, we provide a way to quantify \vepsi given samples.

If \hsim reaches a state where, 
for each matching candidate $i\in \matchingset$,
candidate $i$ has \vepsi-deviation, and 
$\veps_i < \veps$ for all $i\in \matchingset$, 
then it is easy to see that the
\blue{\grec} holds for the matching candidates. 
That is, in such a state, if \hsim output the \viztypes
corresponding to the matching candidates, they would look
similar to the true \viztypes. \blue{In the following sections, we show that
$\veps_i$-deviation can also be used to achieve \gsep.}

\techreport{\topic{\blue{Notation for Round-Specific Quantities}}
\blue{In the following subsections, we use
the superscript ``$\Delta$'' to indicate quantities corresponding
to samples taken during a particular round of \hsim stage 2, such
as $\{\vrip\}$ and $\{\tauip\}$. In particular, these quantities
are {\em completely independent} of samples taken during previous rounds.}}

\subsubsection{Deviation-Bounds Impl\blue{y} \blue{Separation}}\label{subsec:deviations-implies-guarantees}
\noindent In order to reason about the separation guarantee,
\blue{we prove a series of lemmas following the structure of reasoning given below:
\begin{denselist}
\item We show that when a carefully chosen set of null hypotheses are
      all false, $\matchingset$ contains valid top-$k$ closest candidates.
\item Next, we show how to use \vepsi-deviation to upper bound the probability
      of rejecting a {\em single} true null hypothesis.
\item Finally, we show how to reject {\em all} null hypotheses while controlling
      the probability of rejecting {\em any} true ones.
\end{denselist}}
\begin{lemma}[\blue{False Nulls Imply Separation}]
\label{lem:falsenulls}
\blue{Consider the set of null hypotheses $\{H_0\ith\}$ defined as follows,
where $s\in\reals^+$:
\[ H_0\ith =
\begin{cases}
\tausi \geq \nonmatchingbound, \text{ for } i\in\matchingset \\
\tausi \leq \matchingbound, \text{ for } i\in\activeset\setminus\matchingset
\end{cases}
\]
When $H_0\ith$ is false for every $i\in\activeset$, then $\matchingset$ is a
set of top-$k$ candidates that is correct with respect to \gsep.}
\end{lemma}
\techreport{\begin{proof}
\blue{When all the null hypotheses are false, then $\tausi < \nonmatchingbound$ for all $i\in\matchingset$,
and $\tausj > \matchingbound$ for all $j\in\nonmatchingset$. This means that
\[ \max_{i\in\matchingset}\tausi - \min_{j\in\nonmatchingset}\tausj < \veps \]
and thus $\matchingset$ is correct with respect to the separation guarantee.}
\end{proof}}
\blue{Intuitively, \Cref{lem:falsenulls} states that when there is some
reference point $s$ such that all of the candidates in $\matchingset$ have their
$\tausi$ smaller than $\matchingbound$, and the rest have their $\tausi$ greater
than $\nonmatchingbound$, then we have our separation guarantee.}

\blue{Next, we show how to compute P-values for a single null hypothesis
of the type given in \Cref{lem:falsenulls}. Below, we use ``$\P_H$'' to denote
the probability of some event when hypothesis $H$ is true.}
\begin{lemma}[\blue{Distance Deviation Testing}]
\label{lem:disttester}
\blue{Let $x\in\reals^+$.
To test the null hypothesis
$H_0\ith: \tausi \geq x$ versus the alternative $H_A\ith: \tausi < x$, we have that, for any $\veps_i>0$,
\[ \Psubarg{H_0\ith}{x-\tauip > \veps_i} \leq \Parg{d(\vrip, \vrsi) > \veps_i} \]
Likewise, for testing $H_0\ith: \tausi \leq x$ versus the alternative $H_A\ith: \tausi > x$, we have
\papertext{$\Psubarg{H_0\ith}{\tauip-x > \veps_i} \leq \Parg{d(\vrip, \vrsi) > \veps_i}$.}
\techreport{\[\Psubarg{H_0\ith}{\tauip-x > \veps_i} \leq \Parg{d(\vrip, \vrsi) > \veps_i}\]}}
\end{lemma}
\techreport{\begin{proof}
\blue{We prove the first case; the second is symmetric.
Suppose candidate $i$ satisfies
$\tausi \geq x$ for some $x\in\reals^+$. Then, if we take $\takenforip$
samples from which we construct the random quantities $\vrip$ and $\tauip$, we have that
\begin{align*}
\Psubarg{H_0\ith}{x-\tauip > \veps_i} & \leq \Parg{\tausi-\tauip > \veps_i} \\
& = \Parg{\norm{\vhrsi - \vhQ} - \norm{\vhQ - \vhrip} > \veps_i} \\
& \leq \Parg{\norm{\vhrsi - \vhrip} > \veps_i} \\
& = \Parg{d(\vrsi, \vrip) > \veps_i}
\end{align*}
Each step follows from the fact that increasing the quantity to the left of the
``$>$'' sign within the probability expression can only increase the probability of the event inside.
The first step follows from the assumption that $\tausi \geq x$, and
the third step follows from the triangle inequality.}
\end{proof}}
\blue{We use \Cref{lem:disttester} in conjunction with \Cref{lem:falsenulls}
by using $s\pm\frac{\veps}{2}$ for the reference $x$ of \Cref{lem:disttester},
for a particular choice of $s$ (discussed in \Cref{subsec:selecting-deviations}).
For example, \Cref{lem:disttester} shows that when we are testing the null hypothesis
for $i\in\matchingset$ that $\tausi \geq \nonmatchingbound$ and we observe $\tauip$ such
that $0 < \veps_i = \nonmatchingbound - \tauip$, we can use (any upper bound of)
$\Parg{d(\vrsi, \vrip) > \veps_i}$ as a P-value for this test. That is, consider a tester
with the following behavior, illustrated pictorially:}
\begin{center}
\blue{\begin{tikzpicture}
    \draw[thin] (0,-2.3) rectangle (8.3,0);
    \draw[|-|,thick] (0.5,-1.0) -- (4.0,-1.0) node (xspot) {} -- (7.0,-1.0) node (tauspot) {} -- (8.0,-1.0);
    \fill[black] (xspot) circle (0.05);
    \draw[] (xspot) node[above=0.7pt] () {$x$};
    \fill[black] (tauspot) circle (0.05);
    \draw[] (tauspot) node[above=0.7pt] () {$\tauip$};
    \draw [<-|,>=stealth,thin] (0.5,-0.8) -- (3.8,-0.8) node[midway,above=5pt] {$H_0\ith: \tausi \leq x$};
    \draw [decorate,decoration={brace,mirror,amplitude=5pt}] (4.1,-1.1) -- (7.0,-1.1) node[midway,below=5pt] {$\veps_i$};
    \node () at (3.5, -1.9) {If $\Parg{d(\vrsi, \vrip) > \veps_i} \leq \deltaupper$, then {\em reject} $H_0\ith$};
\end{tikzpicture}}
\end{center}
\blue{In the above picture, the tester assumes that $\tausi$ is smaller than $x$,
but it observes a value $\tauip$ that exceeds $x$ by $\veps_i$.
When the true value $\tausi\leq x$ for any reference $x$, then the observed
statistic $\tauip$ will only be $\veps_i$ or larger than $x$ (and vice-versa) when
the reconstruction $\vrip$ is also bad, in the sense that $\Parg{d(\vrsi, \vrip) > \veps_i}$
is very small.}
\blue{If the above tester rejects $H_0\ith$ when $\Parg{d(\vrsi, \vrip) > \veps_i} \leq \deltaupper$,
then \Cref{lem:disttester} says that it
is guaranteed to reject a true null hypothesis with probability at most $\deltaupper$.
We discuss how to compute an upper bound on
$\Parg{d(\vrsi, \vrip) > \veps_i}$ in \Cref{subsec:confidence}.}

\blue{Finally, notice that \Cref{lem:disttester} provides a test which
controls the type 1 error of an individual $H_0\ith$, but we only know
that the separation guarantee holds for $i\in\matchingset$
when {\em all} the hypotheses $\{H_0\ith\}$ are false. Thus, the algorithm
requires a way to control the type 1 error of a procedure that decides
whether to reject every $H_0\ith$ simultaneously. In the next lemma, we
give such a tester which controls the error for any upper bound $\deltaupper$.
\papertext{Our tester is essentially the intersection-union
method~\cite{casella2002statistical} formulated in terms of P-values.}}

\begin{lemma}[\blue{Simultaneous Rejection}]
\label{lem:simultaneous}
\blue{Consider any set of null hypotheses $\{H_0\ith\}$,
and consider a set of P-values $\{\delta_i\}$
associated with these hypotheses. The tester given by
\[
\text{Decision}=
\begin{cases}
\text{reject every $H_0\ith$,} & \text{when $\max\limits_{i} \delta_i \leq \deltaupper$} \\
\text{reject no $H_0\ith$,} & \text{otherwise}
\end{cases}
\]
rejects $\geq 1$ true null hypotheses with probability $\leq \deltaupper$.}
\end{lemma}
\techreport{\begin{proof}
\blue{Consider the set of true null hypotheses and call it $\{H_0\tth\}$ --- suppose there are
$T\geq 1$ in total (if $T=0$, we have nothing to prove), and index them using
$t$ from 1 to $T$. Then
\begin{align*}
\Parg{\exists t : \text{reject } H_0\tth} &= \Parg{\forall t : \text{reject } H_0\tth} \\
&= \, \quad \prod_{t=1}^T \Parg{\text{reject } H_0\tth \,\bigm\vert\, \text{reject } H_0^{(1,\ldots,t-1)}} \\
&= \delta_1 \prod_{t=2}^T \Parg{\text{reject } H_0\tth \,\bigm\vert\, \text{reject } H_0^{(1,\ldots,t-1)}} \\
&\leq \delta_1 \cdot 1 \\
&\leq \deltaupper
\end{align*}
The first step follows since null hypotheses are only rejected when they are all rejected.
The second to last step follows since probabilities are at most 1, and the last step
follows since the tester only rejects when all the P-values are at most $\deltaupper$,
including $\delta_1$.}
\end{proof}}
\techreport{\topic{\blue{Discussion of \Cref{lem:simultaneous}}}
\blue{At first glance, the multiple hypothesis tester given in \Cref{lem:simultaneous},
which compares all P-values to the same $\deltaupper$,
seems to be even more powerful than a Holm-Bonferroni tester,
which compares P-values to various fractions of $\deltaupper$. In fact, although
based on similar ideas, they are not comparable: a Holm-Bonferroni tester may
allow for rejection of a subset of the null hypotheses, wheres the tester of
\Cref{lem:simultaneous} is ``all or nothing''. In fact, the tester of
\Cref{lem:simultaneous} is essentially the union-intersection
method formulated in terms of P-values; see~\cite{casella2002statistical} for
details.}}

\subsubsection{Selecting \blue{Each Round's Tests}}\label{subsec:selecting-deviations}

\techreport{\begin{figure}[t]
 \begin{center}
  \includegraphics[width=.45\textwidth]{\figs/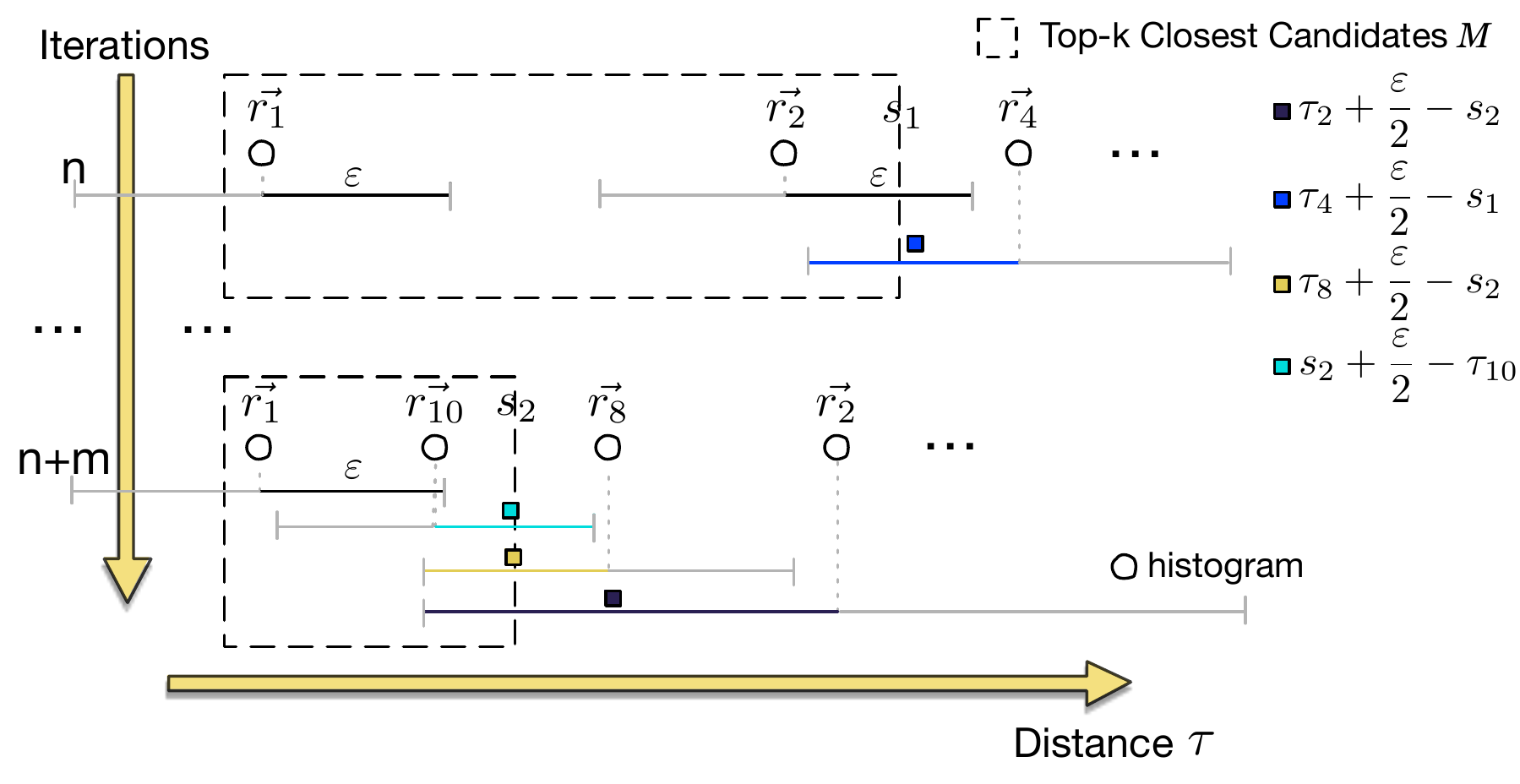}
  \vspace{-10pt}
  \caption{Illustration of \hsim choosing the split point $s$ when
testing whether the separation and reconstruction guarantees hold.} 
 \label{fig:fmeps}
 \end{center}
 \vspace{-1em}
\end{figure}}

\blue{Each round of \hsim stage 2 constructs a family of tests to perform
whose family-wise error probability is at most $\deltaupper$. At round $t$ (starting from $t=1$),
$\deltaupper$ is chosen to be $\frac{\delta/3}{2^t}$, so that the error probability
across {\em all} rounds is at most $\sum_{t\geq1}\frac{\delta/3}{2^t} = \frac{\delta}{3}$
via a union bound (see~\Cref{lem:stage2correct} for details).}

There is still one degree of freedom: namely, how to choose
the split point $s$ \blue{used for the null hypotheses in \Cref{lem:falsenulls}.
In line~\ref{hsim:line:splitchoice}, it is chosen to be
$s \gets \frac{1}{2}(\max\limits_{i\in\matchingset} \tau_i  + \min\limits_{j\in\activeset\setminus\matchingset} \tau_j)$.
The intuition for this choice is as follows. Although the quantities $\vrip$ and $\tauip$
are generated from fresh samples in each round of \hsim stage 2, the quantities $\vri$ and $\taui$
are generated from samples taken across {\em all} rounds of \hsim stage 2. As such, as rounds
progress (i.e., if the testing procedure fails to simultaneously reject multiple times),
the estimates $\vri$
and $\taui$ become closer to $\vrsi$ and $\tausi$, the set $\matchingset$ becomes more
likely to coincide with $\matchingset^*$, and the null hypotheses $\{H_0\ith\}$
chosen become less likely to be true {\em provided} an $s$ chosen somewhere in
$[\max_{i\in\matchingset} \tau_i,
\min_{j\in\activeset\setminus\matchingset}\tau_j]$, since values in this interval
are likely to correctly separate $\matchingset^*$
and $\activeset\setminus\matchingset^*$ as more and more samples are taken.}
In the interest of simplicity, we simply choose the midpoint
halfway between the furthest candidate in  $\matchingset$
and the closest candidate in $\activeset\setminus\matchingset$.
\techreport{For example, at iteration $n$
in \Cref{fig:fmeps}, $s$ lies halfway between candidates $\vr_2$ and $\vr_4$.}
\blue{In practice,
we observe that $\max_{i\in\matchingset}\taui$ and $\min_{j\in\nonmatchingset}\tauj$
are typically very close to each other,
so that the algorithm is not very sensitive to the choice of $s$\papertext{.}\techreport{,
so long as it falls between $\matchingset$ and $\nonmatchingset$.}}
\techreport{
	
\blue{\Cref{fig:fmeps} illustrates this choice of $s$ and the $\{H_0\ith\}$ on our toy example.
As in \Cref{fig:fmtopk}, the boundary of $\matchingset$ is represented by the dashed box.
The split point $s$ is located at the rightmost boundary of the dashed box.}}\papertext{Once $s$ is chosen, the}
\techreport{The} $\{\veps_j\}$ (i.e., the amounts by which the $\{\taujp\}$ deviate from $s\pm\frac{\veps}{2}$)
determine the P-values associated with the $\{H_0\ith\}$ which ultimately determine
whether \hsim stage 2 can terminate, as we discuss more in the next section.

\subsubsection{Deviation-Bounds Given Samples} \label{subsec:confidence}
The previous section provides us a way to
\blue{check whether the rankings induced by the empirical distances $\{\taui\}$
are correct with high probability. This was facilitated via a test
which measures our ``surprise'' for measuring $\{\tauip\}$ if the
current estimate $\matchingset$ is not correct with respect to \gsep,
which in turn used a test for how likely some candidate's
$d(\vrsi, \vrip)$ is greater than some threshold $\veps_i$ after
taking $\takenfori$ samples.}
We now provide a theorem that allows us to infer, given the samples
taken for a given candidate, how to relate \vepsi with
the probability \deli with which the candidate can fail
to respect its deviation-bound \vepsi. \blue{The bound seems
to be known to the theoretical computer science community
as a ``folklore fact''~\cite{ilias};
we give a proof\papertext{~\cite{techreport}} for the sake of completeness.}
Our proof relies on repeated application of the method of bounded
differences~\cite{mcdiarmid1989method} in order to exploit some special structure
in the $\ell_1$ distance metric.
\techreport{The bound developed is {\em information-theoretically
optimal}; that is, it takes asymptotically the fewest samples required to guarantee
that an empirical distribution estimated from the samples will be no further than
\vepsi from the true distribution.}

\vspace{-5pt}
\begin{restatable}{theorem}{thmrec}
 \label{thm:reconstruction}
 Suppose we have taken $n_i$ samples \red{with replacement} for some candidate $i$'s \viztype,
 resulting in the empirical estimate $\vri$. Then $\vri$
 has \vepsi-deviation with probability greater than $1-\delta_i$ for
 $\veps_i = \sqrt{\frac{2}{n_i}\left(\mvx\log2 + \log\frac{1}{\delta_i}\right)}$.
 That is, with probability $> 1-\delta_i$, we have: $\norm{\vhri - \vhrsi}_1 < \veps_i$.
\end{restatable}
\red{In fact, this theorem also holds if we sample without replacement; we return to this
point in \Cref{sec:arch}.}
\techreport{
\vspace{-5pt}
\begin{proof}
For $j \in [\mvx]$, we use $r_j$ to denote the number of occurrences of attribute
value $j$ among the $n_i$ samples, and the normalized count $\hrj$ is our
estimate of $\hrsj$, the true proportion of tuples having value $j$ for
attribute $X$. Note that we have omitted the candidate subscript $i$ for clarity.

We need to introduce some machinery. Consider functions of the form
$ f : [\mvx] \rightarrow \left\{+1, -1\right\}.$
Let $\{f_m\}$ be the set of all such functions, where $m\in [2^\mvx]$,
since there are $2^\mvx$ such functions. For any $m\in [2^\mvx]$, consider
the random variable
\[ Y_m = \sum_{j=1}^\mvx f_m(j)(\hrj - \hrsj) \]
By linearity of expectation, it's clear that $\Earg{Y_m} = 0$, since $f_m(j)$ is
constant and $\Earg{\hrj} = \hrsj$ for each $j$. Since each $\hrj$ is a function of the samples
taken $\{s_k : 1\leq k\leq n_i\}$, each $Y_m$ is likewise uniquely determined from samples, so we can write
$Y_m = g_m(s_1,\ldots,s_{n_i})$, where each sample $s_k$ is a random variable distributed according to $s_k \sim \vhrs$.
Note that the function $g_m$ satisfies the Lipschitz property
\[ |g_m(s_1,\ldots,s_k,\ldots,s_{n_i}) - g_m(s_1,\ldots,s'_k,\ldots,s_{n_i})| \leq \frac{2}{n_i} \]
for any $j \in |\mvx|$ and $s_1,\ldots,s_{n_i}$. For example, this will occur with equality
if $f_m(s_k) = -f_m(s'_k)$; that is, if $f_m$ assigns opposite signs to $s_k$ and $s'_k$,
then changing this single sample moves $1/n_i$ of the empirical mass in such a way that it
does not get canceled out. We may therefore apply the method of bounded differences~\cite{mcdiarmid1989method}
to yield the following McDiarmid inequality\red{---a generalization of the standard Hoeffding's inequality}:
\[ \Parg{Y_m \geq \Earg{Y_m} + \veps_i} \leq \exp{-\veps_i^2n_i / 2} \]
Recalling that $\Earg{Y_m} = 0$, this actually says that
\[ \Parg{Y_m \geq \veps_i} \leq \exp{-\veps_i^2n_i / 2} \]
This holds for any $m \in [2^\mvx]$. Union bounding over all such $m$,
we have that
\[ \Parg{\exists m :  Y_m \geq \veps_i} \leq 2^\mvx \exp{-\veps_i^2n_i / 2} \]
If this does not happen (i.e., for every $Y_m$, we have $Y_m < \veps_i$),
then we have that $\norm{\vhri - \vhrsi}_1 < \veps_i$, 
since for any attribute value $j$, $|\hrj - \hrsj| = \max_{t_j\in\{+1,-1\}} t_j(\hrj - \hrsj)$.
But if $Y_m < \veps_i$ for all $m$, this means that we must have some $m$ such that
\[ \veps_i > \sum_j f_m(j)(\hrj - \hrsj) = \sum_j |\hrj - \hrsj| = \norm{\vhri - \vhrsi}_1 \]
As such $\Parg{\exists m : Y_m \geq \veps_i}$ is an upper bound on $\Parg{\norm{\vhri - \vhrsi}_1 \geq \veps_i}$.
The desired result follows from noting that
\begin{align*}
& \delta_i \leq 2^\mvx \exp{-\veps_i^2n_i / 2} \\
\Longleftrightarrow \,\, &
\veps_i \leq \sqrt{\frac{2}{n_i}\left(\mvx\log2 + \log\frac{1}{\delta_i}\right)} \qed
\end{align*}
\end{proof}
}

\topic{Optimality of the bound in \Cref{thm:reconstruction}}
If we solve for $n_i$ in \Cref{thm:reconstruction}, we see that
\blue{we must} have
$ n_i = \frac{\mvx\log4 + 2\log(1/\delta_i)}{\veps_i^2}. $
That is, $\Omega\left(\frac{\mvx}{\veps_i^2}\right)$ samples are necessary
guarantee that the empirical discrete distribution $\vhri$ is no further than
$\veps_i$ from the true discrete distribution $\vhrsi$, with high probability.
This matches the information theoretical lower bound
noted in prior work~\cite{BatuTestingDistributionsAreClose,chan2014optimal,daskalakis2013learning,waggoner2015p}.

\techreport{\topic{\blue{Generating P-values from \Cref{thm:reconstruction}}}
\blue{We use the above bound to generate P-values for testing the null
hypotheses in \Cref{lem:falsenulls}. From the discussion in
that lemma, a tester which rejects $H_0\ith$ for $i\in M$ when it observes
$s+\frac{\veps}{2}-\tauip > \veps_i$, for fixed $\veps_i$, has a type 1 error
bounded above by $\delta_i = 2^\mvx\exp{-\veps_i^2\takenfori/2}$. Since we want
to bound the type 1 error rate by an amount $\deltaupper$, this induces a particular $\veps_i$
against which we can compare $s+\frac{\veps}{2}-\tauip$, but because $\delta_i$ and $\veps_i$
are monotonically related, we can take
\[ \delta_i = 2^\mvx\exp{-(s+\frac{\veps}{2}-\tauip)^2/2} \]
and compare with $\deltaupper$ directly, allowing us to use this $\delta_i$ as a P-value
for use with the tester in \Cref{lem:simultaneous}.}}

\subsubsection{Stage 2 Correctness}
\techreport{\blue{We can now show correctness of \hsim stage 2.}}
\papertext{\blue{We now formally state the correctness of \hsim stage 2~\cite{techreport}.}}
\begin{lemma}[Stage 2 Correctness]
 \label{lem:stage2correct}
 \blue{After \hsim stage 2 completes, each candidate
 $i\in\matchingset$,
 satisfies $\tausi - \tausj \leq \veps$ for every
 $j\in\activeset\setminus\matchingset$
 with probability greater than $1-\frac{\delta}{3}$.}
\end{lemma}
\techreport{\begin{proof}
\blue{First, show that if \hsim stage 2 terminates after iteration $t$, then the
probability of an error is at most $\frac{\delta/3}{2^t}$. Next, show that the
probability of an error after terminating at {\em any} iteration is at most
$\frac{\delta}{3}$ by union bounding over iterations.}

\blue{If stage 2 terminates at iteration $t$, then the probability of rejecting
one or more null hypotheses is at most $\frac{\delta/3}{2^t}$ by
\Cref{lem:simultaneous} and by \Cref{thm:reconstruction}.
Each $H_0\ith$ for $i\in\matchingset$ says that
$\tausi > s+\frac{\veps}{2}$, and each $H_0\jth$ for $j\in\activeset\setminus\matchingset$
says that $\tausi < s-\frac{\veps}{2}$ -- if all of these are false, then by \Cref{lem:falsenulls}
we have that $\matchingset$ and $\nonmatchingset$ induce
a separation of the candidates that is correct with respect to \gsep, so the
only way an error {\em could} occur is if one or more nulls are true. We just
established that the probability of rejecting one or more true nulls at
iteration $t$ is at most $\frac{\delta/3}{2^t}$, which means that the probability
of an incorrect separation between $\matchingset$ and $\nonmatchingset$
is also at most $\frac{\delta/3}{2^t}$.}

\blue{Finally, by union bounding over iterations, we have that
\begin{align*}
\Parg{\cup_{t\geq 1}\text{mistake at iteration $t$}} &\leq \sum_{t\geq 1}\Parg{\text{mistake at iteration $t$}} \\
&<\sum_{t\geq 1}\frac{\delta/3}{2^t} \ = \ \delta/3
\end{align*}
Thus, when stage 2 terminates, $\matchingset$
is correct (with respect to \gsep) with probability
greater than $1-\frac{\delta}{3}$}
\end{proof}}

\subsection{Stage 3 and Overall Proof of Correctness}\label{subsec:overall-proof}

\blue{Stage 3 of \hsim,
discussed in our overall proof of correctness,
consists of taking samples from
each candidate in $\matchingset$ to ensure they all have
$\veps$-deviation with high probability
(using \Cref{thm:reconstruction}).}
\red{This proof \techreport{is given next, and proceeds in \blue{four} steps:}
\papertext{can be found in~\cite{techreport}; it proceeds in \blue{four} steps:}}
\begin{denselist}
\item \red{Step 1: \blue{\hsim stage 1 incorrectly prunes one or more candidates}
meeting the selectivity threshold $\minsel$ \blue{with probability at most} $\frac{\delta}{\blue{3}}$ (\Cref{lem:stage1correct}).}
\item \red{Step 2: The probability that stage 2 incorrectly \blue{(with respect to \gsep)
separates $\matchingset$ and $\activeset\setminus\matchingset$}
is \blue{at most} $\frac{\delta}{\blue{3}}$}.
\item \blue{Step 3: The probability that the set of candidates $\matchingset$ violates \grec
after stage 3 runs is at most $\frac{\delta}{3}$.}
\item \red{Step \blue{4}: The union bound over \blue{any of} these bad events occurring
\blue{gives an overall error probability of at most $\delta$.}}
\end{denselist}
\begin{theorem}
 \label{thm:correctness}
 \red{The $k$ \viztypes returned by \Cref{alg:basic}
 satisfy \guarantees
 with probability greater than $1-\delta$.}
\end{theorem}
\vspace{-5pt}
\techreport{\begin{proof}
\red{From \Cref{lem:stage1correct}, the probability that high-selectivity candidates were
pruned during stage 1 is upper bounded by $\frac{\delta}{\blue{3}}$.}
\blue{From \Cref{lem:stage2correct}, the probability that the algorithm chooses $\matchingset$
such that there exists some $i\in\matchingset$ and $j\in\matchingset^*\setminus\matchingset$ with $\tausi - \tausj > \veps$
is at most $\frac{\delta}{3}$. Union bounding over these events, the probability of either occurring
is at most $\frac{2\delta}{3}$. Since \gsep cannot be violated when neither of these events occur,
the algorithm violates this guarantee also with probability at most $\frac{2\delta}{3}$.
Finally, using \Cref{thm:reconstruction}, \hsim stage 3 \cref{hsim:line:sample-stage3}
takes a number of samples for each candidate $i\in\matchingset$
such that the probability that a given candidate fails to
be reconstructed with error $\veps$ or less (that is, $d(\vri, \vrsi) > \veps$) is at most $\frac{\delta}{3k}$.
Union bounding over all candidates in $\matchingset$, and noting that $|\matchingset|=k$, the probability that
one or more candidates does not have \vepsi-deviation is at most $\frac{\delta}{3}$. Union bounding with the
upper bound on the probability that \gsep is violated, the probability that either \gsep or \grec is violated
is at most $\frac{2\delta}{3} + \frac{\delta}{3} = \delta$, and we are done.}
\end{proof}}

\topic{Computational Complexity}
\blue{Stage 1 of \Cref{alg:basic} shares computation between candidates
when computing P-values induced by the hypergeometric distribution,
and thus makes at most $\max_{i\in \vz}\takenfori$ calls to evaluate
a hypergeometric pdf (we use Boost's implementation~\cite{boostdist});
this can be done in $\bigo{\max_{i\in\vz}\takenfori}$. To facilitate
the sharing, stage 1 requires sorting the candidates in
increasing order of
$\takenfori$, which is $\bigo{\mvz\cdot\log{\mvz}}$.}
\blue{Next, each iteration of \hsim stage 2
requires computing distance estimates
$\taui$ and $\tauip$ for every $i\in A$, which runs in time
$\bigo{|\activeset|\cdot\mvx}$. Each iteration of stage 2 further
uses a sort of candidates in $\activeset$ by $\taui$ to determine
$\matchingset$ and $s$, which is
$\bigo{|\activeset|\cdot\log{|\activeset|}}$.
\hsim stage 2 almost always terminates within 4 or 5
iterations in practice. Overall, we observe that the computation required is
inexpensive compared to the cost of I/O, even for data stored
in-memory.}

\densesection{The {\large \fm} System}

\label{sec:arch}

This section describes \fm, which implements
the \hsim algorithm. 
We start by presenting the high-level components of
\fm. We then describe the challenges we faced while implementing
\fm and describe how the components interact to alleviate those
challenges, while still satisfying Guarantees~\gsep and~\grec.
\red{While design choices presented in
this section are heuristics with practicality in mind,
the algorithm implemented is still theoretically rigorous,
with results satisfying our probabilistic guarantees.
In the following, each time we describe a 
heuristic, we will clearly point it out as such.
}

\begin{figure}[!t]
\vspace{-10pt}
\begin{center}
    \includegraphics[width=.45\textwidth]{\figs/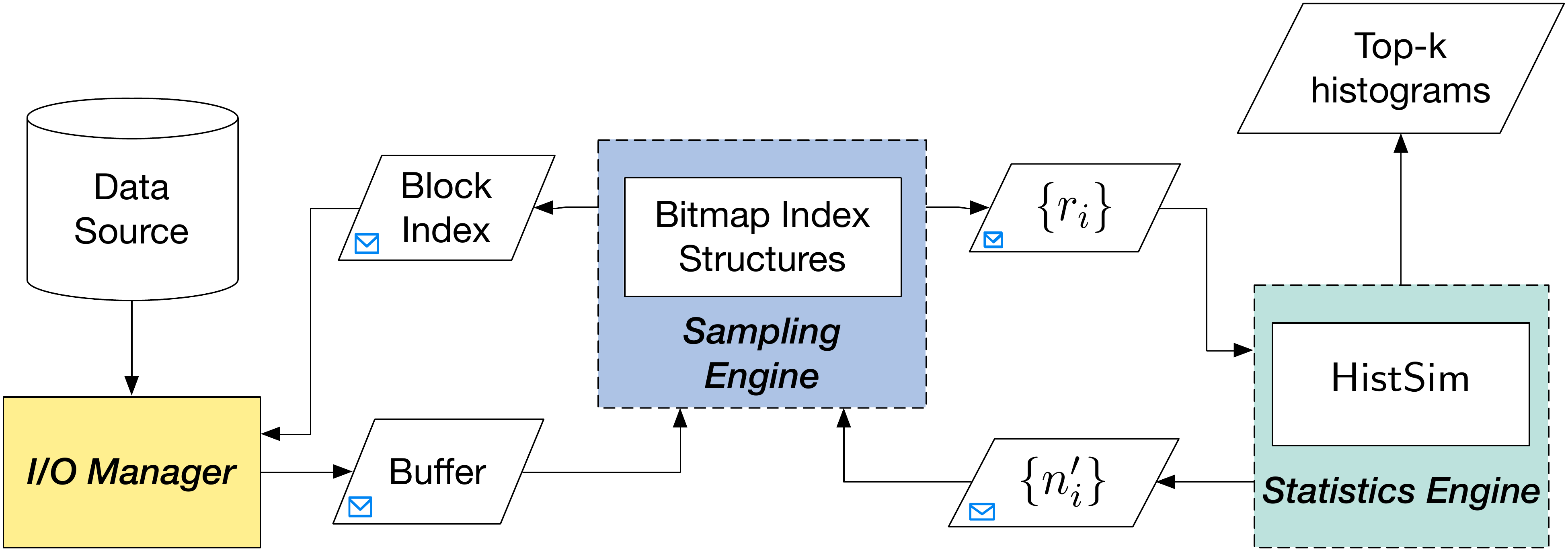}
\end{center}
\vspace{-10pt}
\caption{\fm system architecture}
\label{fig:arch}
\end{figure}

\subsection{{\large \fm} Components}
\fm has three key components:
the I/O Manager, the Sampling Engine, and the Statistics engine.
We describe each of them in turn; \Cref{fig:arch}
provides an architecture diagram---we will revisit 
the interactions within the diagram at the end of the section. 

\topic{I/O Manager}
In \fm, requests for I/O are serviced at the granularity of {\em blocks}.
The I/O manager simply services requests for blocks 
in a synchronous fashion.
Given the location of some block, 
it synchronously \smackoutres{reads}{processes} the block at that location.

\topic{Sampling Engine}
The sampling engine is responsible 
for deciding which blocks to sample. 
It uses
bitmap index structures (described below) in order to determine the types of
samples located at a given block. 
Given the current state of the system,
it prioritizes certain candidates over others for sampling.

\topic{Statistics Engine} 
The statistics engine implements most of the logic in the \hsim algorithm.
The only substantial difference between the actual code and the pseudocode
presented in \Cref{alg:basic} is that the statistics engine does
not actually perform any sampling, instead \blue{leaving this responsibility
to the sampling engine.}
The reason for separating these components
will be made clear later on.


\topic{Bitmap Index Structures}
\fm runs on top of a bitmap-based sampling system \blue{used for
sampling on-demand,}
as in prior work~\cite{alabi2016pfunk,kim2014needletail,Kim2015,rahman2016ve}.
These papers have demonstrated that bitmap indexes~\cite{chan1998bitmap}
are effective in supporting sampling for 
incremental or early termination of visualization
generation. 
Within \fm, bitmap indexes help 
the sampling engine determine whether a given 
block contains samples for a given candidate.
For each attribute $A$, and each attribute value $A_v$, 
we store a bitmap,
where a `0' at position $p$ indicates that the corresponding block at position $p$
contains no tuples with attribute value $A_v$, and a `1' indicates that block $p$
contains one or more tuples with attribute value $A_v$. 
Candidate visualizations
are generated by attribute values\papertext{,}
\techreport{(or a predicate of \texttt{AND}s and \texttt{OR}s over attribute values; see \Cref{sec:extensions}),}
so these bitmaps allow the sampling engine to rapidly test whether a block contains
tuples for a given candidate \viztype.
Bitmaps are amenable to significant compression~\cite{wu2001compressed,wu2008breaking},
and since we are further only requiring a 
single bit per block per attribute value,
our storage requirements are orders-of-magnitude cheaper
than past work that requires a bit per tuple~\cite{alabi2016pfunk,Kim2015,rahman2016ve}.
\techreport{\red{Notice also that our techniques also apply 
for continuous candidate attributes; please see \Cref{sec:extensions}
for details.}}

\subsection{Implementation Challenges}

So far, we have designed \hsim without worrying about
how sampling actually takes place, with an implicit assumption
that there is no overhead to taking samples randomly across
various candidates. 
While implementing \hsim within \fm, 
we faced several non-trivial challenges,
outlined below:


\begin{denselist}
\item {\bf Challenge 1: Random sampling at odds with performance
    characteristics of storage media.} The cost to fetch data is
    locality-dependent when dealing with real storage devices.  
    Even if the data is stored in-memory, tuples (i.e., samples)
    that are spatially closer to a given tuple may be 
    cheaper to fetch, since they may already be present
    in CPU cache.
\item \red{{\bf Challenge 2: Deciding how many samples to take between rounds of \hsim}.
    The \hsim
    algorithm does not specify how many samples to taken in between rounds of stage 2;
    it is agnostic to this choice, with correctness unaffected. If the algorithm takes many samples, it
    may spend more time on I/O than is necessary to terminate with a guarantee. If the algorithm does
    not take enough samples, the statistical test on
    \cref{hsim:line:multitest-stage2} will probably not reject across many rounds, decaying
    $\deltaupper$ and making it progressively
    more difficult to get enough samples to meet \blue{stage 2's} termination criterion.}
\item {\bf Challenge 3: Non-uniform cost/benefit of different candidates.}
    Tuples for some candidates can be over-represented in the data and
    therefore take less time to sample compared to underrepresented candidates. At the same time,
    the benefit of sampling tuples corresponding to different candidate
    \viztypes is non-uniform: for example, those \viztypes which are ``far''
    from the target distribution are less useful (in terms of getting \hsim to
    terminate quickly) than those for which \hsim chooses small values for
    $\veps_i$.
\item {\bf Challenge 4: Assessing benefit to candidates depends on data seen so
    far.} The ``best'' choice of which tuples to sample for
    getting \hsim to terminate quickly can be most accurately estimated from
    {\em all} the data seen so far, including the most recent data. However,
    computing this estimate after processing every tuple and blocking I/O until the
    ``best'' decision can be made is prohibitively expensive.
\end{denselist}

\noindent We now describe our approaches to tackling these three challenges.

\subsubsection*{Challenge 1: Randomness via Data Layout}
To maximize performance benefits from locality, 
we randomly permute the tuples of our
dataset as a preprocessing step, and to ``sample'' we may then simply perform
a linear scan of the shuffled data starting from any point.
\red{This matches the assumptions of stage 1 of \hsim, which
requires samples to be taken without replacement.
Although the theory
we developed in~\Cref{sec:algorithm} for \hsim stage 2 was for sampling with-replacement,
as noted in~\cite{hoeffding1963probability,bardenet2015concentration}, it still holds now that we are
sampling without replacement, 
as concentration
results developed for the with-replacement regime may
be transferred automatically to the without-replacement regime.}
This approach of randomly permuting upfront 
is not new, and is adopted by other approximate
query processing systems~\cite{wu2010continuous,qin2014pf,zeng2015g}.

\subsubsection*{Challenge 2: Deciding Samples to Take Between Rounds}
\red{The \hsim algorithm leaves the number of samples to take during
\blue{a given round of stage 2 lines~\ref{hsim:line:sample-stage2}}
unspecified; its correctness is guaranteed
regardless of how this choice is made. This choice offers a tradeoff: take too
many samples, and the system will spend a lot of time unnecessarily on I/O; take too few,
and the algorithm will never terminate, since the ``difficulty'' of the test increases
with each round, as we set $\deltaupper\gets\deltaupper/2$.}

\papertext{\red{To combat this challenge, we employ a simple heuristic;
the full description of which may be found in~\cite{techreport}.
In brief, our sampling policy is informed by the statistical test
on lines~\ref{hsim:line:multitest-stage2-start}--\ref{hsim:line:multitest-stage2-end}
\blue{--- for each candidate $i$, we attempt to choose a
number of samples to take
$\totakefori$} that will cause \blue{this} test to reject.
We accomplish this by ``inverting the bound'' of \Cref{thm:reconstruction}.
We emphasize that this dependency does not
\blue{compromise the correctness of our results}
thanks to the union bound between rounds of stage 2
(see Theorem~\ref{thm:correctness}) \blue{and since
each round's test uses fresh samples to compute
the test statistics $\{\tauip\}$}.}}

\techreport{\red{To combat this challenge, we employ a simple heuristic.
To estimate the
number of samples we need to take for candidate \blue{$i$},
\blue{we assume that $\taui = \tausi$, so that we need to learn
$\vrip$ to within $\veps_i'$ of $\vrsi$ for a given round's statistical test to
successfully reject, where
$\veps'_i = s+\frac{\veps}{2}-\tau_i$ for $i\in\matchingset$
and $\veps'_i = \tau_i - (s-\frac{\veps}{2})$ for
$i\in\activeset\setminus\matchingset$. (Recall that we use \vepsi-deviation
to upper bound the P-values.)
For this setting of
$\{\veps'_i\}$, we thus choose to take samples
for each candidate}
by solving for $\takenfori$ in the bound of \Cref{thm:reconstruction}. 
This yields}
\begin{equation}
\label{eqn:stage2esttotake}
\totakefori = 2\left(\mvx\log{2} - \log{\deltaupper}\right) / \left(\veps'_i\right)^2
\end{equation}
\red{Each round of stage 2 of our \fm implementation of \hsim thus continues to take samples until
$\takenforip \geq \totakefori$ for every candidate $i$. It then performs
the multiple hypothesis test on
lines~\ref{hsim:line:multitest-stage2-start}--\ref{hsim:line:multitest-stage2-end}.
If it rejects,
the algorithm terminates and the system gives
the output to the user; otherwise,
it once again estimates each $\totakefori$ using \Cref{eqn:stage2esttotake}
\blue{(plugging in $\{\veps'_i\}$ from updated $\{\taui\}$)} and repeats.}}

\subsubsection*{Challenge 3: Block Choice Policies}

\red{Deciding which blocks to read during stage 1 of \hsim is simple since
we are only trying to detect low-selectivity candidates --- in this case
we just scan each block sequentially. Deciding which blocks to read
during stage 2 of \hsim is more difficult due to} the non-uniform
cost (i.e., time) and benefit of samples for 
each candidate \viztype. If either cost or benefit were uniform across candidates, matters would
be simplified significantly: if cost were uniform, we could simply read in the blocks
with the most beneficial candidates; if benefit were uniform, we could simply read in
the lowest cost blocks (for example, those closest spatially to the current read position). 
To address these concerns, we developed
a simple policy which we found worked well in practice for getting \hsim to terminate quickly.

\topic{\anyactive block selection policy}
Recall that \red{the end of} each iteration of \red{stage 2 of} \hsim \smackoutres{computes a 
set of deviations $\{\delta_i\}$, where each $\delta_i$ is
an upper bound on the probability that candidate $i$ fails to have \vepsi-deviation.}{estimates
the number of samples $\{\totakefori\}$ necessary from each candidate so that the next
iteration is more likely to terminate.}
Note that
\smackoutres{if every $\delta_i$ satisfied $\delta_i \leq \frac{\delta}{\mvz}$,}{if
each candidate satisfied $\takenfori = \totakefori$ at the time \hsim performed
the test for termination and {\em before} it computed the \{$\totakefori$\}},
then \hsim would be
in a state where it can safely terminate. 
Those candidates for whom \smackoutres{$\delta_i > \frac{\delta}{\mvz}$}{$\takenfori < \totakefori$}
we dub {\em active candidates}, and we employ a very simple block selection policy, dubbed the
\anyactive block selection policy, which is to {\em only read blocks which contain at least one tuple
corresponding to some active candidate}.
The bitmap indexes employed by \fm allow it to rapidly test whether a block contains
tuples for a given candidate visualization, and thus to rapidly
apply the \anyactive block selection policy. 
Overall, our approach is as follows: we read blocks in sequence,
and if blocks satisfy our \anyactive criterion, then
we read all of the tuples in that block, else, we skip that block.
\blue{We discuss how to make this approach performant below.}

\techreport{A naive variant of this policy is presented
in \Cref{alg:anyactive_bad}, for which we describe improvements below.}

\hidden{\topic{Policy 2: locality-aware block selection}
This policy mainly benefits reads from traditional hard disks, where the cost to
fetch data is linear in the distance from the current position, and then constant.
This policy is as follows: if the next block to be read is within a distance of $b$
from the current position, simply read all blocks between the current position and $b$,
inclusive. We found that a good setting for $b$ on our hard disk was 1000. For in-memory
experiments, we set $b=0$.

In \fm, our sampling engine combines the above two policies for a {\em locality-aware \anyactive}
block selection policy.}

\subsubsection*{Challenge 4: Asynchronous Block Selection}

\techreport{\begin{table*}
\begin{tabular}{c|c}
\begin{minipage}{0.5\textwidth}

\def\cand{\texttt{cand}\xspace}
\def\index{\texttt{index\_lookup}\xspace}
\def\read{\texttt{:read}\xspace}
\def\skip{\texttt{:skip}\xspace}

\begin{algo}[t]
 \captionof{algocf}{Naive \anyactive block processing}\label{alg:anyactive_bad}
\SetKwInOut{Input}{Input}
\SetKwInOut{Output}{Output}
{\small

\Input{\red{unpruned} candidate set $A$, block index $i$}
\Output{A value indicating whether to \read or \skip block $i$}
\BlankLine
\BlankLine
 \For{\em \red{each active} \cand $\in A$}{
    \tcp{cache inefficient index lookup}
    \tcp{evicts bits from previous candidate's bitmap index}
    \If{\em \cand.\index(i)}{
	  \Return{\em \read}\;
    }
 }
 \Return{\em \skip}\;
}

\BlankLine
\BlankLine
\BlankLine
\BlankLine
\BlankLine
\BlankLine
\BlankLine
\BlankLine
\vspace{2pt}
\end{algo}

\end{minipage}
&
\begin{minipage}{0.5\textwidth}

\def\cand{\texttt{cand}\xspace}
\def\mark{\texttt{mark}\xspace}
\def\start{\texttt{start}\xspace}
\def\index{\texttt{index\_lookup}\xspace}
\def\read{\texttt{:read}\xspace}
\def\skip{\texttt{:skip}\xspace}

\begin{algo}[t]
 \captionof{algocf}{\anyactive block selection with \lookahead}\label{alg:anyactive_good}
\SetKwInOut{Input}{Input}
\SetKwInOut{Output}{Output}
{\small

\Input{\lookahead amount, \start block, \red{unpruned} candidate set $A$}
\Output{An array \mark indicating whether to \read or \skip blocks}
\BlankLine
\tcp{Initialization} \mark[i] $\gets$ \skip for $0\leq i<$ \lookahead\;
\For{\em \red{each active} \cand $\in A$}{
 \For{$0\leq i <$ {\em \lookahead}}{
    \uIf{\em \mark[i] == \read}{
	\texttt{continue}\;
    }\ElseIf{\em \cand.\index(\start + i)}{
	\mark[i] $\gets$ \read\;
    }
 }
}
\Return{\em \mark}
}

\end{algo}

\end{minipage}
\end{tabular}
\end{table*}}

\begin{figure}[t]
\begin{center}
    \includegraphics[width=.4\textwidth]{\figs/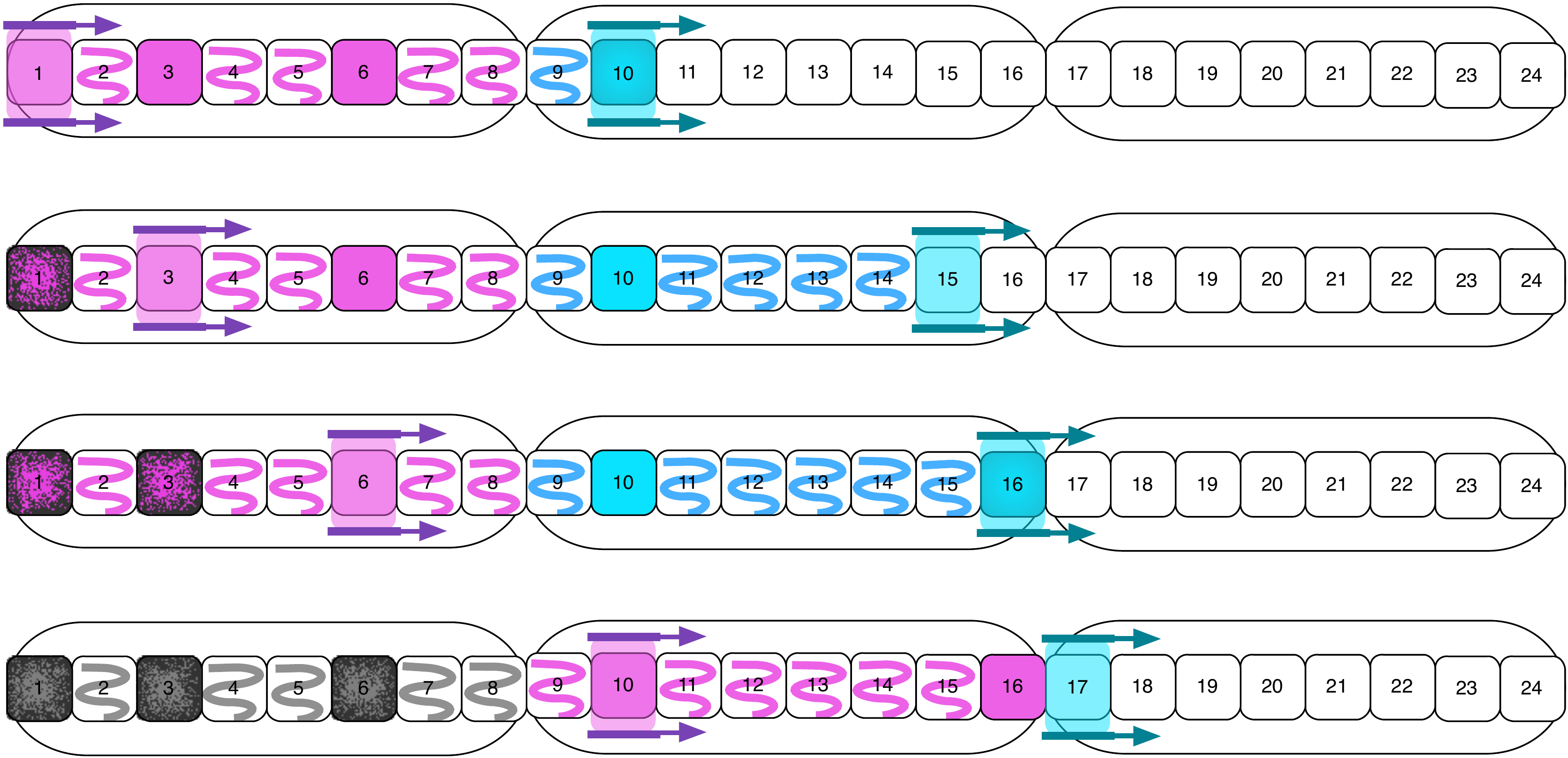}
\end{center}
\vspace{-10pt}
\caption{
While the I/O manager processes magenta blocks, the sampling engine selects
blue blocks ahead of time, using \lookahead.
Blocks with solid color = read, blocks with squiggles = skip.}
\label{fig:lookahead}
\end{figure}

From the previous discussion, the sampling engine employs an \anyactive
block selection policy when deciding which blocks to process. \red{Ideally,
the \smackoutres{$\{\delta_i\}$}{$\{\takenfori\}$ and $\{\totakefori\}$ (number
of samples taken for candidate $i$ and estimated number of samples needed for candidate $i$, respectively)}
used to assign active status to candidates should be
computed from the freshest possible \smackoutres{data}{counts} available to
\smackoutres{\hsim}{the sampling engine}. \red{That is, in
an ideal setting, each candidate's active status would be updated immediately after
each block is read, and the potentially new active status should be used for making
decisions about immediately subsequent blocks.} Unfortunately,
this requirement is at odds with real system characteristics. Employing it
exactly implies \smackoutres{running an iteration of \hsim to compute fresh $\{\delta_i\}$
after every block is processed. This causes two intertwined problems.}{leaving
the I/O manager idle while the sampling engine determines whether
each block should be read or skipped.}}
\red{To \smackoutres{do so}{prevent this issue}, we relax the requirement that the
sampling thread employ \anyactive with the freshest \smackoutres{$\{\delta_i\}$}{$\{\takenfori\}$} available
to it. Instead, given \red{the current $\{\takenfori\}$ and} fresh set of
\smackoutres{$\{\delta_i\}$}{$\{\totakefori\}$},
it \red{precomputes the active status for each candidate and}} ``looks ahead'',
marking an entire batch of blocks for either reading or skipping, and communicates
this with the I/O manager. The batch size, or the \lookahead amount, is a \blue{system} parameter,
and offers a trade-off between freshness of \smackoutres{$\{\delta_i\}$}{active states} used
for \anyactive and degree to which the I/O manager must idle while waiting for instructions on
which block to read next.
We evaluate the impact of this parameter in our experimental section.
The \lookahead process
is depicted in \Cref{fig:lookahead} for a value of \lookahead $=8$. While the I/O
manager processes a previously marked batch of magenta-colored \lookahead blocks,
the sampling engine's \lookahead thread marks the next batch in blue. 
It waits to mark the next batch until the I/O manager ``catches up''.

Employing \lookahead allows us to prevent two bottlenecks. First,
the sampling engine need not wait for \smackoutres{the freshest $\{\delta_i\}$
from the statistics engine}{each candidate's active status to update after a block is read
before moving on to the next block, effectively decoupling it from the I/O manager.}

The second bottleneck prevented by \lookahead is more subtle.
\papertext{A detailed description, with comparisons and pseudocode
can be found in our technical report~\cite{techreport}.
Here, we simply provide a high-level idea.}\techreport{To illustrate it, consider the pseudocode in \Cref{alg:anyactive_bad},
implementing the \anyactive block policy.}
The \anyactive block policy algorithm works by considering each candidate in turn, 
and querying a bitmap
index for that candidate to determine whether the current block contains tuples
corresponding to that candidate. Querying a bitmap actually brings in surrounding
bits into the cache of the CPU performing the query, and evicts whatever was
previously in the cache line. If blocks are processed individually, then only
a single bit in the bitmap is used each time a portion is brought into cache.
This is quite wasteful and turns out to hurt performance significantly as we
will see in the experiments.
Instead, applying \anyactive selection to \lookahead-size chunks instead of individual
blocks is a better approach.
\techreport{This simply adds an extra inner loop to the procedure shown in
\Cref{alg:anyactive_bad} (depicted in
\Cref{alg:anyactive_good}).
This approach has much better cache
performance, since it uses an entire cache-line's worth of bits while employing
\anyactive.}
\papertext{This approach has much better cache
performance, since it allows an entire cache-line's worth of bits to be used.}

\papertext{\blue{We verify the benefits of these optimizations in our experiments.}}
\techreport{We verify in our experiments that these optimizations allow \fm to terminate
more quickly via \anyactive block selection with {\em fresh-enough} \smackoutres{$\{\delta_i\}$}{active states}
without significantly slowing any single component of the system.}

\subsection{System Architecture}
\fm is implemented within a few
thousand lines of C\texttt{++}. It uses \texttt{pthreads}~\cite{pthreads}
for its threading implementation.
\smackoutres{We use \fm in a row-store mode in order
to obviate any benefits from column-stored data and better test
the components described in this section.}{\fm uses a column-oriented
storage engine, as is common for analytics tasks.}
We can now complete our description of \Cref{fig:arch}. When the I/O manager receives
a request for a block at a particular block index from the sampling
engine (via the ``block index'' message), it eventually returns a
buffer containing the data at this block to the sampling engine (via the ``buffer'' message).
\red{Once the I/O phase of stage 1 or 2 of \hsim completes, the sampling engine sends the current
per-group counts for each candidate, $\{\vri\}$, to the statistics engine.}
After running a \smackoutres{\hsim iteration}{test for whether to move to stage 2 (performed in stage 1)
or to terminate (performed in stage 2)}, the statistics engine \red{either} posts a message of updated
\smackoutres{$\{\delta_i\}$}{$n'$ (in stage 1) or $\{\totakefori\}$ (stage 2)}
\smackoutres{used by the sampling engine for block selection}{that
the sampling engine uses to determine when to complete the I/O phase of each \hsim stage,
as well as how to perform block selection during stage 2}.

\densesection{Experimental  Evaluation}
\label{sec:experiments}


\begin{table}[!t]
\scriptsize
\vspace{-5pt}
\begin{center}
\begin{tabular}{ | c || c|c|c|c|} \hline
  \textbf{Dataset} & Size & \#Tuples & \#Attributes & Replications\\ \hline \hline
  \flights & 32 GiB & \numflightstuples  & 7  & $5\times$ \\ \hline
  \taxi    & 36 GiB & 679 million & 7  & $4\times$ \\ \hline
  \police  & 34 GiB & 448 million & 10 & $72\times$ \\ \hline
\end{tabular}
\vspace{-10pt}
\caption{Descriptions of Datasets}
\label{tab:datasets}
\end{center}
\vspace{-10pt}
\end{table}

\begin{table*}[!t]
\scriptsize
\vspace{-10pt}
\begin{center}
\begin{tabular}{ | c | c || c|c|c|c|} \hline
  \textbf{Dataset} & \textbf{Query} & Z \ ($\mvz$) & X \ ($\mvx$) & \preck & target \\ \hline \hline
  \flights & $q_1$ & Origin (347) & DepartureHour (24) & 10 & Chicago ORD \\ \hline
           & $q_2$ & Origin (347) & DepartureHour (24) & 10 & Appleton ATW \\ \hline
           & $q_3$ & Origin (347) & DayOfWeek (7) & 5 & [0.25, 0.125, 0.125, 0.125, 0.125, 0.125, 0.125] \\ \hline
           & $q_4$ & Origin (347) & Dest (351) & 10 & closest candidate to uniform \\ \hline \hline
  \taxi & $q_1$ & Location (7641) & HourOfDay (24) & 10 & closest candidate to uniform \\ \hline
        & $q_2$ & Location (7641) & MonthOfYear (12) & 10 & closest candidate to uniform \\ \hline \hline
  \police & $q_1$ & RoadID (210) & ContrabandFound (2) & 10 & closest candidate to uniform \\ \hline
          & $q_2$ & RoadID (210) & OfficerRace (5) & 10 & closest candidate to uniform \\ \hline
          & $q_3$ & Violation (2110) & DriverGender (2) & 5 & closest candidate to uniform \\ \hline
\end{tabular}
\vspace{-10pt}
\caption{Summary of queries}
\label{tab:queries}
\end{center}
\vspace{-20pt}
\end{table*}

The goal of our experimental evaluation is to test
the accuracy and runtime of \fm 
against other approximate and exact approaches
on a diverse set of real datasets and queries.
Furthermore, we want to validate the 
design decisions that we made for \fm in \Cref{sec:arch} 
and evaluate their impact.


\subsection{Datasets and Queries}

We evaluate \fm on publicly available real-world datasets 
summarized in \Cref{tab:datasets} --- flight
records~\cite{flights}, taxi trips~\cite{taxi}, and police road stops~\cite{police}.
The replication value indicates how many times each dataset was
replicated to create a larger dataset.
In preprocessing these datasets, we eliminated
rows with ``N/A'' or erroneous values for \smackoutres{certain columns;
and to ensure that there is an adequate number of samples
across candidates, we pruned tuples with attribute values that appear
fewer than 2000 times in the data (such tuples constituted fewer
than 1\% of the data, for every dataset).}{any column appearing
in one or more of our queries.}
\papertext{Details on the datasets and attributes can be found
in our technical report.}

\techreport{
\topic{\flights Dataset}
Our \flights dataset,
representing delays measured for flights at more than 350 U.S. airports from
1987 up to 2008, is available at~\cite{flights};
we used 7 attributes (for origin / destination airports, departure / arrival delays,
day of week, day of month, and departure hour).

\topic{\taxi Dataset}
Our \taxi dataset summarizes all Yellow Cab
trips in New York in 2013~\cite{taxi}. The subset of data we used corresponds with the
urls ending in ``yellow\_tripdata\_2013''
in the file \texttt{raw\_\-data\_urls.txt}.
We extracted some time-based discrete attributes, two attributes
based on passenger count, and one attribute based on area, for 7 columns total. 
In particular, the ``Location'' attribute was generated by binning the pickup location into regions of 0.01 longitude
by 0.01 latitude. \techreport{As with our \flights data, we discarded rows with missing values, as well as
rows with outlier longitude or latitude values (which did not correspond to real locations).}
\red{The taxi data stressed our algorithm's ability to deal with low-selectivity candidates,
since more than 3000 candidates have fewer than 10 total datapoints.}

\topic{\police Dataset}
Our \police dataset
summarizes more than 8 million police road stops in Washington state~\cite{police}. 
We extracted attributes for county, two gender attributes,
two race attributes, road number, violation type, stop outcome, whether a search was conducted, and
whether contraband was found, for 10 attributes total. 
}

\techreport{\topic{Queries and Query Format}}
We evaluate several queries on each dataset, whose templates are summarized in \Cref{tab:queries}.
We had four queries on \flights, \flights-q1-q4, two on \taxi, \taxi-q1-q2, and 
three on \police, \police-q1-q3.
\techreport{For simplicity, in all queries
we test, the x-axis is generated by grouping over a single attribute
(denoted by ``X'' in \Cref{tab:queries}),
and the different candidates are likewise generated
by grouping over a single (different) attribute (signified by ``Z'').}
For each query, the visual target was chosen to correspond with the closest
distribution (under $\ell_1$) to uniform, out of all \viztypes generated via
the query's template, except for q1, q2, and q3 of \flights.
Our queries spanned a number of interesting dimensions:
{\em (i) frequently-appearing top-$k$ candidates:}
\flights-q1, \police-q1 and q2,
{\em (ii) rarely-appearing top-$k$ candidates:}
\flights-q2 and q3,
{\em (iii) high-cardinality candidate attribute $Z$:}
\taxi-q1 and q2 ($\mvz = 7641$), \police-q3 ($\mvz=2110$), and
{\em (iv): high-cardinality grouping attribute $X$:}
\flights-q4 ($\mvx=351$).
\red{The taxi queries in particular stressed our algorithm's ability to deal with low-selectivity candidates,
since more than 3000 locations have fewer than 10 total datapoints.}


\begin{table}[!t]
\scriptsize
\begin{center}
\blue{\begin{tabular}{ | c || c || c|c|c|} \hline
  \textbf{Query} &            & \multicolumn{3}{c|}{Avg Speedup over \scan{} (raw time in (s))} \\ \hline \hline
                 & \scan (s) & \scanmatch & \syncmatch & \fm \\ \hline \hline
F-q1 & $12.26$ & $27.74\times$~~(0.44) & $25.53\times$~~(0.48) & $\mathbf{37.52\times}$~~(0.33) \\ \hline
F-q2 & $12.29$ & $3.17\times$~~(3.87) & $2.73\times$~~(4.51) & $\mathbf{10.11\times}$~~(1.21) \\ \hline
F-q3 & $11.62$ & $4.76\times$~~(2.44) & $3.14\times$~~(3.70) & $\mathbf{8.72\times}$~~(1.33) \\ \hline
F-q4 & $13.97$ & $5.93\times$~~(2.36) & $5.76\times$~~(2.43) & $\mathbf{8.15\times}$~~(1.71) \\ \hline
T-q1 & $13.09$ & $4.89\times$~~(2.68) & $0.32\times$~~(40.95) & $\mathbf{15.93\times}$~~(0.82) \\ \hline
T-q2 & $13.09$ & $6.48\times$~~(2.02) & $0.37\times$~~(35.60) & $\mathbf{17.38\times}$~~(0.75) \\ \hline
P-q1 & $8.57$ & $5.72\times$~~(1.50) & $5.14\times$~~(1.67) & $\mathbf{13.34\times}$~~(0.64) \\ \hline
P-q2 & $8.49$ & $14.31\times$~~(0.59) & $15.48\times$~~(0.55) & $\mathbf{36.11\times}$~~(0.24) \\ \hline
P-q3 & $8.65$ & $9.25\times$~~(0.93) & $1.53\times$~~(5.66) & $\mathbf{33.26\times}$~~(0.26) \\ \hline
\end{tabular}}
\vspace{-10pt}
\caption{\blue{Summary of average query speedups and latencies}}
\label{tab:latencies}
\end{center}
\vspace{-5pt}
\end{table}

\subsection{Experimental Setup}

\topic{Approaches}
We compare \fm against a number of less sophisticated 
approaches
that provide the same guarantee as \fm. 
\smackoutres{Except for \scan,
they are all parametrized by}{All approaches are parametrized by a minimum selectivity
threshold $\minsel$, and all approaches except \scan are additionally
parametrized by}
$\veps$ and $\delta$ and satisfy
\guarantees with probability greater than $1-\delta$.
\begin{denselist}
\item \syncmatch($\veps,\delta,\red{\minsel}$).
This approach uses \fm, but the \anyactive block
selection policy is applied without \lookahead, 
synchronously and for each individual block.
{\em By comparing this method with \fm, we quantify
how much benefit we may ascribe to the \lookahead technique.}
\item \scanmatch($\veps,\delta,\red{\minsel}$).
This approach uses \fm, but without
the \anyactive block selection policy.
Instead, no blocks are pruned: 
it scans through each block in a 
sequential fashion until
the statistics engine reports that \hsim's termination criterion holds.
{\em By comparing this with \syncmatch, we quantify
how much benefit we may ascribe to \anyactive block selection.}
\hidden{\item \slowmatch($\veps,\delta$).
This approach uses \scanmatch, but with a
termination criterion that requires more samples.
Specifically, the statistics engine runs a variant of \hsim that
generates $1-\frac{\delta}{\mvz}$ confidence intervals around each candidate, and
then tests whether (a) any of the top-$k$ estimated candidates have intervals wider
than $\veps$, or (b) any of the top-$k$ overlap any of the non-top-$k$ by an amount
more than $\veps$. If either (a) or (b) hold, the algorithm does not terminate.
Note that this is equivalent to running \hsim until $\max_i\{\delta_i\} \leq \frac{\delta}{\mvz}$
instead of the more intelligent termination criterion $\sum_i \delta_i \leq \delta$.
{\em By comparing this method with \scanmatch, we quantify
how much benefit we may ascribe to \hsim's intelligent termination criterion.}}
\item \scan(\red{$\minsel$}).
This approach is a simple heap scan over the entire dataset and always returns correct results,
trivially satisfying \guarantees. \red{It exactly prunes candidates with selectivity below $\minsel$.}
{\em By comparing \scan with our above approximate
approaches, we quantify how much benefit we may ascribe to the use of approximation.}
\end{denselist}

\topic{Environment}
Experiments were run on single Intel Xeon E5-2630 node with 125 GiB of RAM and with 8 physical cores
(16 logical) each running at 2.40 GHz, although we use at most \red{2} logical cores
to run \fm components.
The Level 1, Level 2, and
Level 3 CPU cache sizes are, respectively: 512 KiB, 2048 KiB, and
20480 KiB. We ran Linux with kernel version 2.6.32.
We report results for data stored in-memory,
since the cost of
main memory has decreased to the point that most interactive workloads
can be performed entirely in-core.
Each run of \fm or any other approximate approach
was started from a random position in the shuffled data.
\smackoutres{We found wall clock time to be
very stable for all approaches, so any times are reported as the average of \smackout{5}{30} runs for all methods.}{We
report both wall clock times and accuracy as the average across 30 runs with identical parameters,
with the exception of \scan{}, whose wall clock times we report as
the average over 5 runs.}
\smackoutres{Experiments related to accuracy had higher variance; we report any accuracy-related
metrics averaged across \smackout{30}{FIXME} runs for all methods (excepting \scan).}{}
Where applicable, we used default settings of \blue{$m=5\cdot10^5$}, $\delta=0.01$, $\veps=\red{0.04}$,
\red{$\minsel=0.0008$}, and \lookahead$=\red{1024}$.
\red{We set the block size for each column to \smackoutres{4 KiB, which matches the sector sizes for
typical storage devices}{600 bytes, which we found to perform well}; our
results are not too sensitive to this choice.}

\subsection{Metrics}

We use several metrics to compare \fm against our baselines
in order to test two hypotheses: one, that \fm does indeed provide
accurate answers, and two, that the system architecture developed in
\Cref{sec:arch} does indeed allow for earlier termination
while satisfying the separation and reconstruction guarantees.

\topic{Wall-Clock Time}
Our primary metric evaluates the end-to-end time
of our approximate approaches that are variants of \fm,
as well as a scan-based baseline.


\topic{Satisfaction of Guarantees~\gsep and~\grec}
Our $\delta$ parameter ($\delta=0.01$), serves as an upper bound on the
probability that either of these guarantees are violated. If this bound
were tight, we would expect to see about one run in every hundred fail
to satisfy our guarantees. We therefore count the number of times our
guarantees are violated
\blue{relative to}
the number of queries performed.

\hidden{\topic{Fraction of $\veps$-deviant \viztypes in output}
\smacke{Discuss how we evaluate based on proportion of \viztypes in output
that look close to truth}

\topic{P@\preck} We evaluate our approximation algorithms in terms of
the proportion of the estimated top-\preck-closest visualizations which
coincide with the actual top-\preck.
Recall that $\matchingset$ is the set of indexes of \viztypes output
by \fm, and that $\matchingset^*$ is the set of \preck indexes whose
corresponding \viztypes are the true closest to some target $\vQ$.
Mathematically,

\[ P@K = \frac{|\matchingset \cap \matchingset^*|}{K} \]

\topic{$\veps$-tolerant P@\preck}
\smacke{variant of P@K that is 1 when separation guarantee satisfied -- fill me in}}

\topic{Total Relative Error in Visual Distance}
In some situations, there may be several candidate \viztypes
that are quite close to the analyst-supplied target, and choosing
any one of them to be among the \preck returned to the analyst would
be a good choice.
We define the
{\em total relative error in visual distance}
(denoted by $\Delta_d$) between the \preck candidates
returned by \fm and the true \preck closest visualizations as:
$\Delta_d(\matchingset, \matchingset^*, \vQ) = \frac{\sum_{i\in\matchingset} d(\vri, \vQ) - 
    \sum_{j^\in \matchingset^*}d(\vrsj, \vQ)}{\sum_{j\in\matchingset^*}d(\vrsj, \vQ)} $
\smackoutres{We always have that $\Delta_d \geq 0$, with smaller values corresponding to
higher-quality results.}{Note that here, $M^*$ is computed by \scan and only considers candidates meeting the
selectivity threshold. Since \fm and our other approximate variants have no recall requirements
with respect to identifying low-selectivity candidates (they only have precision requirements),
it is possible for $\Delta_d < 0$.}

%


\subsection{Empirical Results}
\begin{figure*}[t]
\vspace{-10pt}
  \begin{center}
    \includegraphics[width=0.8\textwidth]{\figs/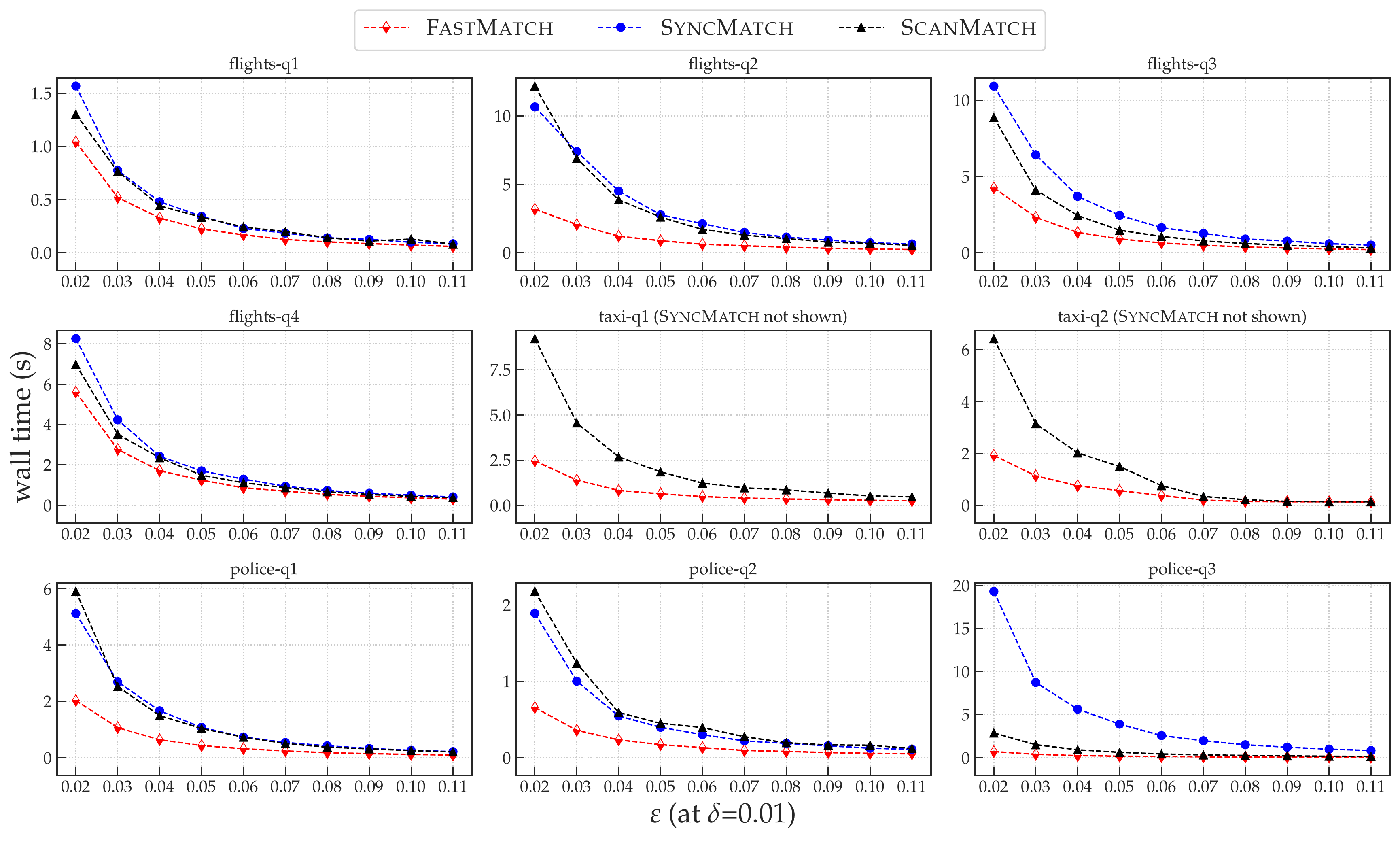}
  \end{center}
  \vspace{-20pt}
  \caption{\blue{Effect of $\veps$ on query latency}}
  \label{fig:time_vs_eps}
  \vspace{-10pt}
\end{figure*}

\begin{figure*}[t]
\vspace{-10pt}
  \begin{center}
    \includegraphics[width=0.8\textwidth]{\figs/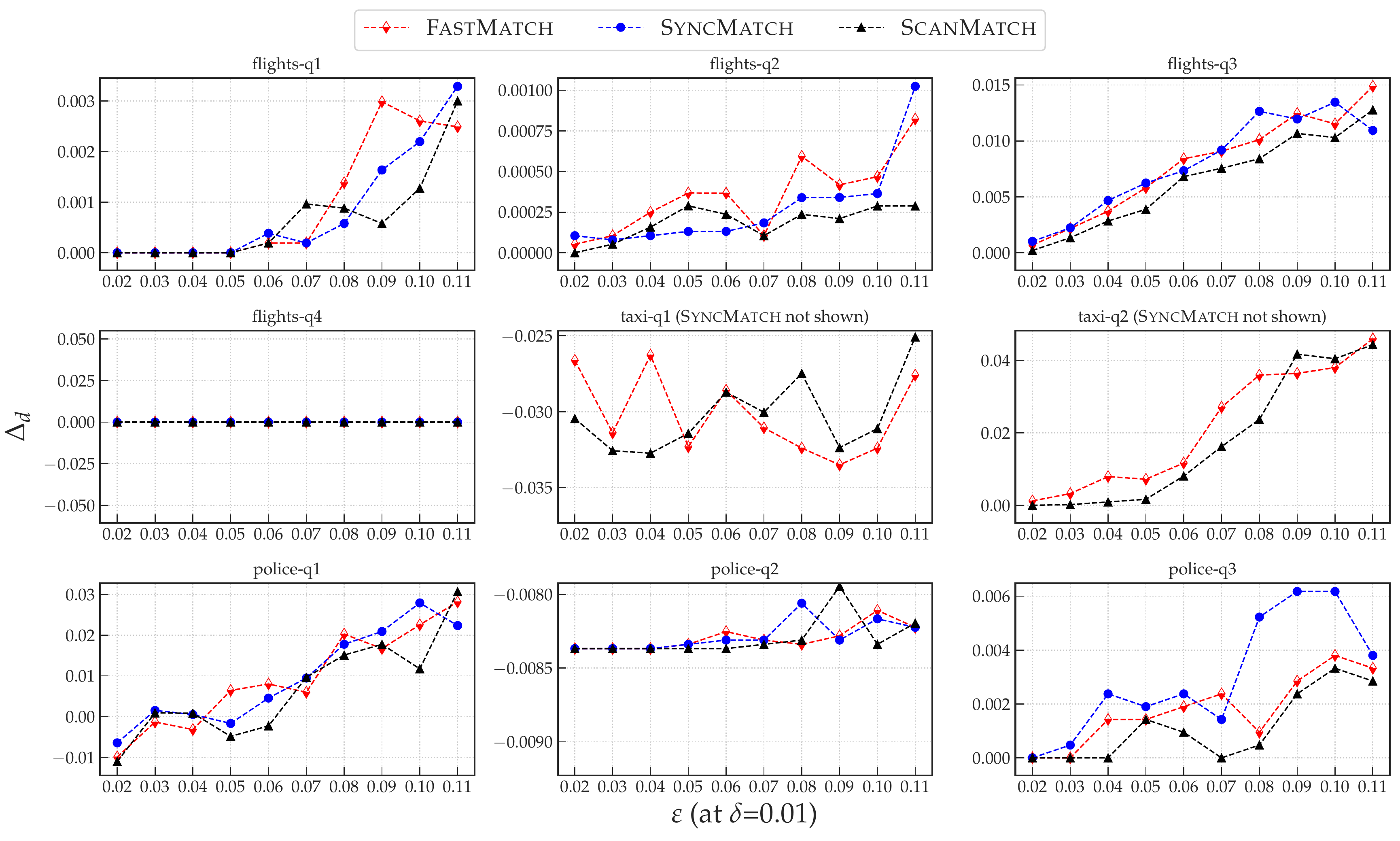}
  \end{center}
  \vspace{-20pt}
  \caption{\blue{Effect of $\veps$ on $\Delta_d$}}
  \label{fig:acc_vs_eps}
\vspace{-10pt}
\end{figure*}
\begin{figure*}[t]
\centering
  \includegraphics[width=0.8\textwidth]{\figs/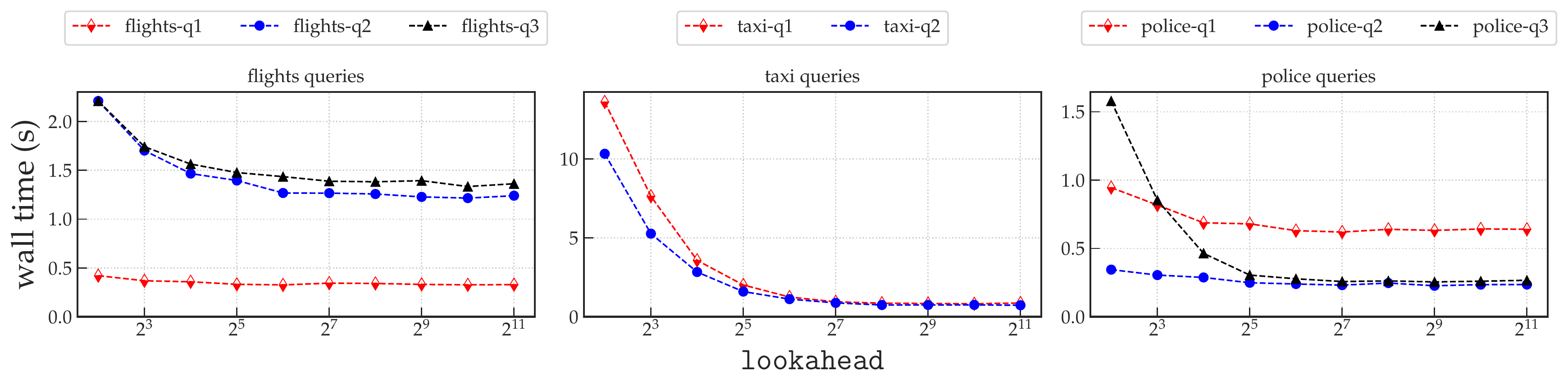}
  \vspace{-15pt}
  \caption{\blue{Effect of varying lookahead}}
  \label{fig:time_vs_lookahead}
  \vspace{-10pt}
\end{figure*}
\techreport{\begin{figure*}[t]
\centering
  \includegraphics[width=0.8\textwidth]{\figs/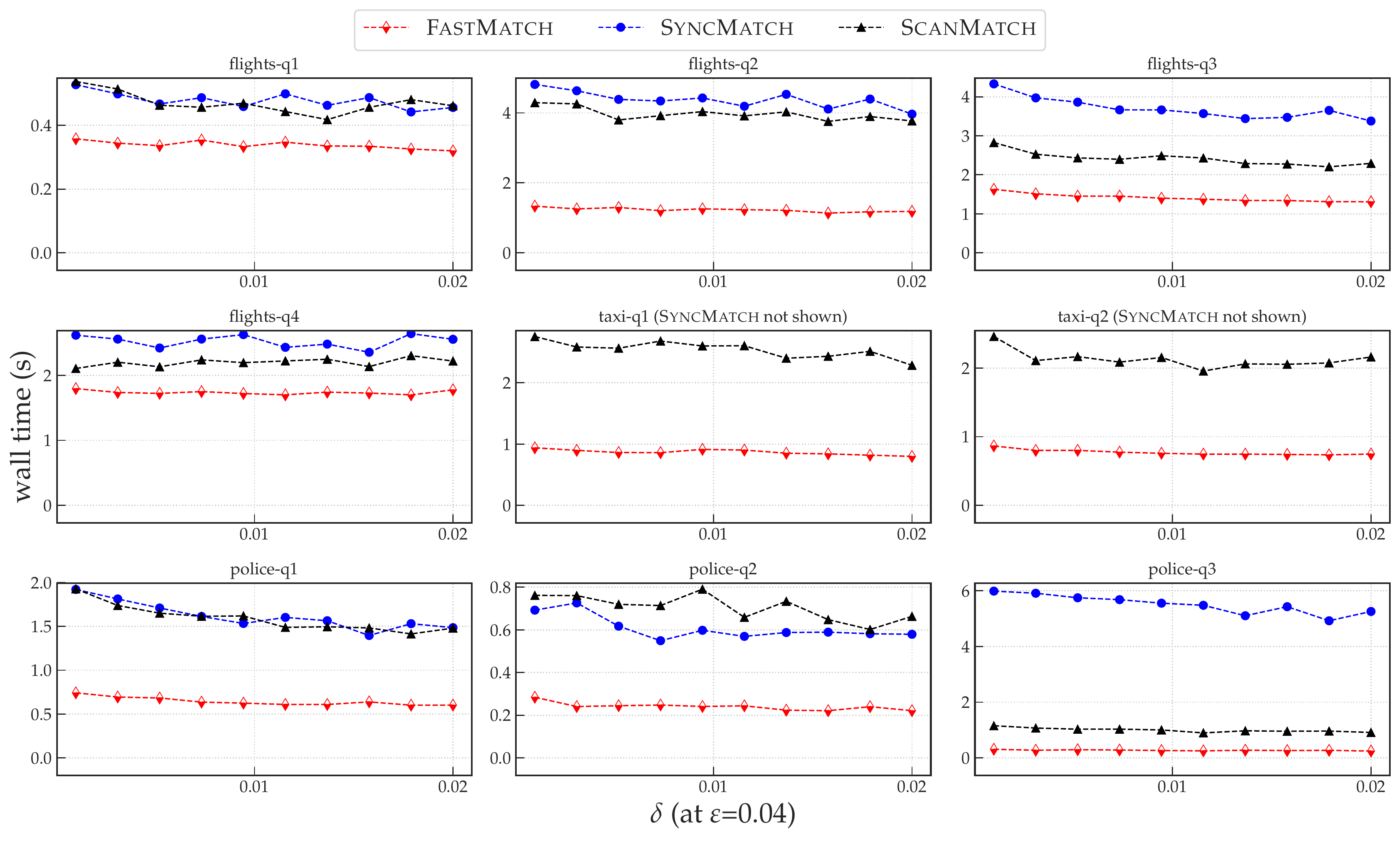}
  \vspace{-15pt}
  \caption{Effect of $\delta$ on wall clock time}
  \label{fig:time_vs_delta}
  \vspace{-10pt}
\end{figure*}}
%
%

\topic{Speedups and Error of \fm versus others}

\frameme{
\noindent {\bf \em Summary.}
All \fm variants we tested show significant speedups over \scan
for at least one query, but only \fm shows consistently excellent performance,
typically beating other approaches and bringing latencies for all queries near
interactive levels; with an overall speedup ranging between 
\blue{$\mathbf{\minspeedup}$ \textbf{and} $\mathbf{\maxspeedup}$} over \scan.
Further, the output of \textbf{\fm and all approximate variants
satisfied \guarantees across all runs for all queries.}
}
Average run times of \fm and other approaches, 
for all queries as well as speedups over \scan,
are summarized in \Cref{tab:latencies}.
We used default settings for all runs.
The reported speedups are the ratio of the average wall time
of \scan with the average wall time of each approach considered.
\scan was generally slower than approximate approaches because
it had to examine all the data. Then, we typically observed that
\red{\smackoutres{\slowmatch took longer than \scanmatch due to its worse termination
criterion, which in turn generally took longer than \syncmatch because
it was not able to employ \anyactive block selection}{\scanmatch and \syncmatch
were pretty evenly matched, with \blue{\scanmatch} usually performing slightly better,
except in some pathological cases where it performed very poorly due to
poor cache usage}. \fm  had
 better performance than \red{either} \syncmatch
\red{or \scanmatch}, thanks to \lookahead \red{paired with \anyactive block selection}.
Overall, we observed that each of \fm's key innovations:
the termination criterion, the block selection, and \lookahead,
all led to substantial performance improvements,
with an overall speedup of up to \smackoutres{{\bf 136$\times$}}{$\mathbf{\blue{\maxspeedup}}$}
over \scan.}

\blue{Queries with high candidate cardinality (\taxi-q*, \police-q3), displayed
particularly interesting performance differences. For these,}
\fm shows greatly improved performance over \scanmatch. 
\red{It also scales much better to the large number of candidates than} \syncmatch, which
performs extremely poorly due to poor cache utilization
and takes around $3\times$
longer than a simple non-approximate \scan. In this case, the \lookahead
technique of \fm is necessary to reap the benefits of \anyactive block selection.


Additionally, we found that the output of \emph{\fm and all approximate variants
satisfied \guarantees across all runs for all queries.}
This suggests that the parameter $\delta$
may be a loose upper bound for the actual failure probability of \fm.

\topic{Effect of varying $\veps$}
\frameme{
\noindent {\bf \em Summary.}
In almost all cases, increasing the tolerance parameter $\veps$ leads to reduced
runtime and accuracy, but
\textbf{on average, $\Delta_d$ was never more than \smackoutres{6\%}{\blue{5\%}}
larger than optimal for any query, even for the largest values of $\veps$ used.}
}
\Cref{fig:time_vs_eps,fig:acc_vs_eps} depict the effect of
varying $\veps$ on the wall clock time and on $\Delta_d$, respectively,
using $\delta=0.01$ and \lookahead = \smackoutres{512}{1024}, averaged over \smackoutres{5 runs for
wall clock time and 30 runs for $\Delta_d$}{30 runs} for each value of $\veps$.
Because of the extremely poor performance of \syncmatch
on the \taxi queries, we omit it from both figures.

In general, as we increased $\veps$, wall clock time decreased and $\Delta_d$
increased. In some cases,
\red{\scanmatch latencies
matched that of \scan until we made $\veps$ large enough.} This
sometimes happened when it needed more refined estimates
of the (relatively infrequent) top-$k$ candidates, which it achieved by scanning most of the data,
picking up lots of superfluous (in terms of achieving safe termination) tuples along the way.


\topic{Effect of varying \lookahead}
\techreport{\frameme{\noindent {\bf \em Summary.}
When the number of candidates $\mvz$ is not large, performance is relatively stable
as \lookahead varies. For large $\mvz$,  more \lookahead helps performance, but is not crucial.}}
For most queries, we found that latency was relatively robust to changes in \lookahead.
\Cref{fig:time_vs_lookahead} depicts this effect.
\blue{The queries with high candidate cardinalities (\taxi-q*, \police-q3) were the exceptions.}
\blue{For these queries,}
larger \lookahead values led to increased utilization at all levels of CPU cache.
\red{Past a certain point, \blue{however,}} the performance gains were minor. 
\red{Overall}, we found the default value of \red{1024} to be acceptable in all circumstances.

\topic{Effect of varying $\delta$}
\papertext{We refer readers to the technical report~\cite{techreport} for a
full discussion (including experimental plots) on the effect of varying $\delta$.}
In general, we found that \red{increasing $\delta$ led to slight decreases in wall
clock time, leaving accuracy (in terms of $\Delta_d$) more or less constant.} We believe this
behavior is inherited from our bound in \Cref{thm:reconstruction}, which is not sensitive
to changes in $\delta$.
\techreport{\Cref{fig:time_vs_delta} shows the effect of varying $\delta$
on wall clock time. For the values of $\delta$ we tried, we did not observe
any meaningful trends in $\Delta_d$ and have omitted the plot.}

\topic{When approximation performs poorly}
\red{In order to achieve the competitive results presented
in this section, the initial pruning of low-selectivity
candidates during stage 1 of \hsim ended up being critical
for good performance. With a selectivity threshold of $\minsel=0$,
\blue{stages 2 and 3} of \hsim \blue{are} forced to consider many extremely rare candidates.
For example, in the taxi queries, nearly half of candidates have fewer than
10 corresponding datapoints. In this case, \scanmatch performs the best (essentially performing a \scan
with a slight amount of additional overhead),
but it (necessarily) fails to take enough samples to establish \guarantees.
\syncmatch and \fm likewise fail to establish guarantees, but additionally have the issue
of being forced to consider many rare candidates while employing \anyactive block
selection, which can slow town query processing by a factor of $100\times$ or more.}


\techreport{\begin{table}[!t]
\small
\begin{center}
\begin{tabular}{ | c || c|c|} \hline
  \textbf{Query} & $\frac{|M^*(\ell_1)\cap M^*(\ell_2)|}{k}$ & Relative distance difference \\ \hline \hline
  \flights-$q_1$ & 0.9 & 0.01 \\ \hline
  \flights-$q_2$ & 0.7 & 0.04 \\ \hline
  \flights-$q_3$ & 0.6 & 0.03 \\ \hline
  \flights-$q_4$ & 0.8 & 0.01 \\ \hline
\end{tabular}
\vspace{-10pt}
\caption{Comparison of top-closest histograms for $\ell_1$ and $\ell_2$}
\label{tab:distcomp}
\end{center}
\vspace{-5pt}
\end{table}}

\newcommand{\reltotaldist}{\frac{\sum_{j\in M^*(\ell_2)}d(\vrj,\vQ) - \sum_{i\in M^*(\ell_1)}d(\vri,\vQ)}{\sum_{i\in M^*(\vri, \vQ)}d(\vri,\vQ)}}

\topic{Comparing results for $\ell_1$ and $\ell_2$ metrics}
\red{So far, we have not validated our choice of distance metric (normalized $\ell_1$);
prior work has shown that normalized $\ell_2$ is suitable
for assessing the ``visual'' similarity of visualizations~\cite{vartak2015s}, so here, we compare our top-k
with the top-k using the normalized $\ell_2$ metric, for the \flights
queries. In brief, we found that the relative difference in the total $\ell_1$ distance of
the top-k using the two metrics never exceeded 4\% for any query, and 
that roughly 75\% of the top-k candidates were common across the two metrics. 
Thus, $\ell_1$ can serve as a suitable replacement for  $\ell_2$, while
further benefiting from the advantages we described in \Cref{sec:problem}.
\papertext{Please see~\cite{techreport} for the full discussion.}
\techreport{Table~\ref{tab:distcomp} summarizes our full results.}
}

%
%
%
%

\densesection{Related Work}
\label{sec:related}

We now briefly cover work that is related to \fm from a number of different areas.

\stitle{Approximate Query Processing (AQP).} 
Offline AQP involves computing 
a set of samples offline, and then using these samples when
queries arrive~e.g.,~\cite{ioannidis1999histogram,
chaudhuri2001overcoming,acharya2000congressional,alon1996space, blinkdb},
with systems like BlinkDB~\cite{blinkdb}
and Aqua~\cite{acharya1999aqua}.
These techniques crucially
rely on the availability of a workload.
On the other hand, online approximate query processing, e.g.,~\cite{Hellerstein1997,hou1989processing,lipton1993efficient}, 
performs sampling on-the-fly, typically using an index to facilitate
the identification of appropriate samples.
Our work falls into the latter category; however, none of the prior
work has addressed a similar problem of identifying relevant visualizations given a query.

\stitle{Top-K or Nearest Neighbor Query Processing.}
\red{
There is a vast body of work on top-k query 
processing~\cite{DBLP:journals/csur/IlyasBS08}.
Most of this work relies on exact answers, as opposed to approximate
answers, and has different objectives.
\techreport{As an example, Bruno et al.~\cite{Bruno2002}
exploit statistics maintained by
a RDBMS in order to quickly find top-$k$ tuples 
matching user-specified
attribute values. }
Some work tries to bridge the gap between
top-k query processing and uncertain 
query processing~\cite{DBLP:conf/icde/SolimanIC07,DBLP:conf/icde/ReDS07,silberstein2006sampling,cohen2008processing,pietracaprina2010mining,chen2002sampling,kriegel2007probabilistic,beskales2008efficient}, 
but does not need to deal with the concerns of where and when to sample
to return answers quickly, but approximately.
Some of this work~\cite{DBLP:conf/icde/SolimanIC07,DBLP:conf/icde/ReDS07,kriegel2007probabilistic,beskales2008efficient} develops efficient algorithms for top-k or nearest neighbors in a uncertain databases setting---here, the sampling is restricted to monte-carlo sampling, which is very different in behavior. 
Silberstein et al.~\cite{silberstein2006sampling} retain samples of past sensor readings
to avoid maintaining joint probability distributions in a sensor network.
Cohen et al.~\cite{cohen2008processing} develops techniques to bound the probability
of a given set of items being part of the top-k.
Pietracaprina et al.~\cite{pietracaprina2010mining} develops sampling schemes tailored
to finding top-k frequent itemsets.
Chen et al.~\cite{chen2002sampling} employ sampling to determine the bounds of 
a minimum bounding rectangle for top-k nearest neighbor queries. 
Zhang et al.~\cite{zhang2017fast} performs top-k similarity search efficiently in a network context.
}

\stitle{Scalable Visualizations.}
There has been some limited work on scalable approximate
visualizations, targeting the generation of a single visualization,
while preserving certain properties~\cite{Kim2015,Park2015,rahman2016ve}.
In our setting, the space of sampling 
is much larger---as a result the problem is more complex. 
Furthermore, the objectives are very different.
Fisher et al.~\cite{Fisher2012} explores the impact of approximate
visualizations on users, adopting an online-aggregation-like~\cite{Hellerstein1997}
scheme. 
As such, these papers show that users are able to interpret
and utilize approximate visualizations correctly.
Some work uses pre-materialization for the purpose of 
displaying visualizations quickly~\cite{kandel2012profiler,DBLP:journals/tvcg/LinsKS13,liu2013immens};
however, these techniques rely on in-memory data cubes.
We covered other work on scalable visualization
via approximation~\cite{SPS,moritz2017trust,jugel2014m4,Park2015,wu2015efficient,vartak2015s} in \Cref{sec:intro}.

\stitle{Histogram Estimation for Query Optimization.}
A number of related papers~\cite{chaudhuri1998random,ioannidis1995balancing,jagadish1998optimal}
are concerned with the problem of sampling for histogram estimation, usually for estimating
attribute value selectivities~\cite{lipton1990practical} and query size estimation
(see~\cite{chen2017two} for a recent example). While some of the theoretical tools
used are similar, the problem is fundamentally
different, in that the aforementioned line of work is concerned with estimating
one histogram per table or view for
query optimization purposes with low error, while we are concerned with comparing
histograms to a specific target.

\stitle{Sublinear Time Algorithms.}
\hsim is related to work on sublinear time algorithms---the most relevant ones~\cite{BatuTestingDistributionsAreClose,chan2014optimal,waggoner2015p}
fall under the setting of distribution learning and analysis of {\em property testers}
for whether distributions are close under $\ell_1$ distance. Although Chan et al.~\cite{chan2014optimal} develop bounds for testing whether distributions
are $\veps$-close in the $\ell_1$ metric, property testers can only say when two distributions
$p$ and $q$ are equal or $\veps$-far, and  cannot handle $\norm{p-q}_1 < \veps$ for $p\ne q$, 
a necessary component of this work.

\densesection{Conclusion\techreport{ and Future Work}}
\label{sec:conclusion}
We developed sampling-based strategies for rapidly identifying the top-$k$
\viztypes that are closest to a target.
We designed a general algorithm, \hsim,
that provides a principled framework to facilitate this search, with theoretical guarantees.
We showed how
the systems-level optimizations present in our \fm architecture are crucial for achieving
near-interactive latencies consistently, leading to speedups ranging from \blue{$\minspeedup$} to
\blue{$\maxspeedup$} over baselines.
\techreport{While this work suggests several possible avenues for further
exploration, we are particularly interested in exploring the
impact of our systems architecture in supporting general interactive analysis.}

\balance
\bibliographystyle{abbrv}
\bibliography{main}
\techreport{\appendix

\section{Extensions}
\label{sec:extensions}
\subsection{Generalizing Problem Description}
\subsubsection{\SUM aggregations}
While we do not consider it explicitly in this paper, in~\cite{SPS}, the authors
describe how to perform \SUM aggregations with $\ell_2$ distributional guarantees
via {\em measure-biased sampling.} Briefly, a measured-biased sample for some attribute
$Y$ involves sampling each tuple $t$ in $T$, where the probability of inclusion in the sample
is proportional $t$'s value of $Y$. \fm can also leverage measure-biased samples in order
to match bar graphs generated via the following types of queries:
\begin{quote}
\normalfont
\small
 SELECT $X$, \SUM(Y) FROM $T$\\
 WHERE $Z=z_i$ GROUP BY $X$
\end{quote}
As in \Cref{def:vizquery}, $Z$ is the candidate attribute and $X$ is the grouping attribute
for the x-axis. One measure-biased sample must be created per measure attribute $Y$ the analyst
is interested in, so if there are $n$ such attributes, we require an additional $n$ complete
passes over the data for preprocessing. When matching bar graphs generated according to the
above template, \fm would simply use the measure-biased sample for $Y$ and pretend as if
it were matching visualizations generated according to \Cref{def:vizquery}; that is, it would
use \COUNT instead of \SUM. There is nothing special about the $\ell_2$ metric used in~\cite{SPS},
and the same techniques may be used by \fm to process queries satisfying \guarantees.

\subsubsection{Candidates based off arbitrary boolean predicates}

In order to support candidates based off boolean predicates such as $Z\fst = z\fst_i \wedge Z\snd = z\snd_j$,
\fm needs a way to estimate the number of active tuples in a block for the purposes of applying
\anyactive block selection. In this case, simple bitmap indexes with one bit per block are not enough.
We may instead opt to use the slightly costlier {\em density maps} from~\cite{needletailoptimal}.
We refer readers to that paper for a description of how to estimate the
number of tuples in a block satisfying an arbitrary boolean predicate. Even if different candidates share
some of the same tuples, our guarantees still hold since \hsim uses \red{a Holm-Bonferroni procedure
to get joint guarantees across different candidates at a given iteration, a method which is agnostic
to any dependency structure between candidates.}

\subsubsection{Multiple attributes in \texttt{GROUP BY} clause}

In the case where the analyst wishes to use multiple attributes $X\fst, X\snd, \ldots, X\nth$ to generate the
support of our histograms generated via \Cref{def:vizquery}, all of the same methods apply, but we estimate
the support $\mvx$ as \[ |V_{X\fst}| \cdot |V_{X\snd}| \cdot \ldots \cdot |V_{X\nth}| \]

This may be an overestimate if two attribute values, say $x\fst_i$ and $x\snd_j$, never occur together.
Our guarantees still hold in this case --- overestimating the size of the support can only make the bound
in \Cref{thm:reconstruction} looser than it could be, which does not cause any correctness issues.

\subsubsection{Handling continuous $X$ attributes via binning}
If the analyst wishes to use a continuous $X$, she must simply provide a set of non-overlapping bin ranges, or
``buckets'' in which to collect tuples. Everything else is still the same. \red{In fact, \flights-q1 and \flights-q2
used this technique, since the DepartureHour attribute was actually a continuous attribute we placed
into 24 bins (although we presented it as a discrete attribute for simplicity).}

\subsubsection{\red{Handling an Unknown Candidate Domain}}
\red{If the candidate domain is unknown at query time, for example if we do not
have any bitmap index structures over the attribute(s) used to generate candidates,
it is still possible to use a variant of our methods. First of all, we may still
employ \scanmatch, creating state for new candidates as they are discovered.
During stage 1 of \hsim, in which rare candidates are pruned, we must also account
for any potential candidates for which \hsim has not yet seen any tuples. In this
case, we may simply add one additional ``dummy'' candidate which matches against
all the tuples for any unseen candidates. We add an additional test to the
Holm-Bonferroni procedure for this dummy candidate --- if the test rejects,
and if $U$ represents the indices of the unseen candidates, then we can
be sure that $\frac{\sum_{j\in U}\Ntotalj}{\Ntotal} < \minsel$, which in
turn implies that $\frac{\Ntotalj}{\Ntotal} < \minsel$ for each $j\in U$.}

\subsubsection{\red{Handling Continuous Candidates}}
\red{If one or more of the attributes used to group candidates is continuous,
then, as in the case of continuous $X$, candidates may be ``grouped'' by
placing different real-values into bins. We can also construct bitmaps for
continuous attributes at some predetermined finest level of granularity
of binning, which can then be used to induce bitmaps for any coarser granularity that may be
needed. Even if the finest granularity available is too coarse to isolate different
candidates, as long as it isolates some subsets of candidates, it may still be useful
for pruning the blocks that need to be considered for \anyactive block selection.
Even if there is no index available, one may still use \scanmatch.}

\subsection{Different Types of Guarantees}
\subsubsection{Allowing Distinct $\veps_1$ and $\veps_2$ for \guarantees}

If the analyst believes one of \guarantees is more important than the other,
she may indicate this by providing separate $\veps_1$ for \gsep and $\veps_2$
for \grec; \hsim generalizes in a very straightforward way in this case.
For example, if \grec is more important than \gsep, the analyst may provide
$\veps_1$ and $\veps_2$ with $\veps_2 < \veps_1$.

\subsubsection{Allowing other distance metrics}

We can extend \hsim to work for any distance metric for which there exists an
analogue to \Cref{thm:reconstruction}. For example, there exist such bounds
for $\ell_2$ distance~\cite{SPS,waggoner2015p}. 

\subsubsection{Allowing a range of $\preck$ in input}
In some cases, the analyst may not care about the exact number of matching
candidates. For example, the analyst may be fine with finding anywhere between
5 and 10 of the closest histograms to a target. In this case, she may specify a
range $[k_1, k_2]$, and \fm may return some number $k\in[k_1,k_2]$ of histograms
matching the target, where $k$ is automatically picked to make it as easy as possible
to satisfy \guarantees. For example, in the case $[k_1, k_2] = [5, 10]$, there may
be a very large separation between the $7$th- and $8$th-closest candidates, in which
case \hsim can automatically choose $k=7$, as this likely provides a small
$\deltaupper$ as soon as possible.
}

\end{document}